\documentclass[useAMS,usenatbib]{mnras}

\usepackage{graphicx}	% Including figure files
\usepackage{amsmath}	% Advanced maths commands
\usepackage{amssymb}	% Extra maths symbols
\usepackage[dvipsnames]{xcolor}

\definecolor{orange}{rgb}{1,0.5,0}

\newcommand{\gsim}{\,\lower2truept\hbox{${>\atop\hbox{\raise4truept\hbox{$\sim$}}}$}\,}

\newcommand{\be}{\begin{equation}}
\newcommand{\ee}{\end{equation}}
\newcommand{\bea}{\begin{eqnarray}}
\newcommand{\eea}{\end{eqnarray}}

\newcommand{\zbar}{\bar{Z}}

\newcommand{\ud}{\mathrm{d}}
\newcommand{\eff}{{\rm{eff}}}

\newcommand{\current}{S}

\def\aap{A\&A}
\def\apj{ApJ}

\def\apjl{ApJ}
\def\mnras{MNRAS}

\def\apjs{ApJS}

\renewcommand{\vec}[1]{ {\bmath #1} } 

\def\ltsima{$\; \buildrel < \over \sim \;$}
\def\simlt{\lower.5ex\hbox{\ltsima}}
\def\gtsima{$\; \buildrel > \over \sim \;$}
\def\simgt{\lower.5ex\hbox{\gtsima}}

\title[Dark Scattering beyond constant w]
{{Structure formation simulations with momentum exchange: alleviating tensions between high-redshift and low-redshift cosmological probes}}
\author[M.~Baldi \& F.~Simpson]{\parbox{\textwidth}{Marco Baldi$^{1,2,3}$, Fergus Simpson$^{4}$}
\\
\\$^{1}$Dipartimento di Fisica e Astronomia, Alma Mater Studiorum Universit\`a di Bologna, viale Berti Pichat, 6/2, I-40127 Bologna, Italy;
\\$^{2}$INAF - Osservatorio Astronomico di Bologna, via Ranzani 1, I-40127 Bologna, Italy;
\\$^{3}$INFN - Sezione di Bologna, viale Berti Pichat 6/2, I-40127 Bologna, Italy;
\\$^{4}$ICC, University of Barcelona (UB-IEEC), Marti i Franques 1, 08028, Barcelona, Spain.}

\def\m@th{\mathsurround=0pt }
\def\eqalign#1{\null\,\vcenter{\openup1\jot \m@th
 \ialign{\strut\hfil$\displaystyle{##}$&$\displaystyle{{}##}$\hfil
 \crcr#1\crcr}}\,}

\begin{document}
%\date{Accepted ???. Received ???; in original form }
\pagerange{\pageref{firstpage}--\pageref{lastpage}} \pubyear{2011}
{\color{red}
\maketitle
}
\label{firstpage}
\begin{abstract}	
\\
 Persisting tensions between the cosmological constraints derived from low-redshift probes and the ones obtained from temperature and polarisation anisotropies of the Cosmic Microwave Background -- {although not yet providing compelling} evidence against the $\Lambda $CDM model -- seem to consistently indicate a slower growth of density perturbations as compared to the predictions of the standard cosmological scenario. Such behavior {is not easily accommodated by the simplest extensions of General Relativity, such as $f(R)$ models, which generically predict an enhanced growth rate}. In the present work we present the outcomes of a suite of large N-body simulations carried out in the context of a cosmological model featuring a non-vanishing scattering cross section between the dark matter and the dark energy fields, for two different parameterisations of the dark energy equation of state. Our results indicate that these Dark Scattering models have very mild effects on 
%the background expansion history of the universe as well as on  
many observables related to large-scale structures formation and evolution, while providing a significant suppression of the amplitude of linear density perturbations and the abundance of massive clusters. {Our simulations therefore confirm that these models offer a promising route to alleviate existing tensions between low-redshift measurements and those of the CMB}.
\end{abstract}


%*****************************************************************************

\section{Introduction}
\label{i} 

%% Introduction to dark energy and dark matter and why they may be coupled
{The physical characteristics of dark matter and dark energy remain poorly constrained.  Proposed candidates for dark matter  include axions, WIMPs, and black holes, while dark energy has been linked to a cosmological constant and scalar fields. And of course, there remains ample scope for either dark matter or dark energy to be described by a fundamentally new form of physics. One fairly well determined feature is their current energy density, and in that respect, these two phenomena appear comparable. This has fuelled speculation that their relationship is not purely gravitational.}

%% Experimental Evidence
A further hint of a connection between the universe's two dominant constituents stems from small but consistent deviations between low redshift measurements of the amplitude of density perturbations, and the extrapolated value based on the amplitude of primary anisotropies in the Cosmic Microwave Background \citep[CMB][]{Planck_2015_XIII}. These low redshift measurements include weak gravitational lensing from CFHTLenS \citep[][]{syspaper}, redshift space distortions induced by the motions of galaxies \citep[][]{BOSSRSD2012, BlakeWigglezRSD, Simpson_etal_2016}, and galaxy clusters \citep[][]{Xrays}, all indicating a slightly lower amplitude of clustering than has been inferred from the Cosmic Microwave Background. Additionally, even lensing of the CMB itself prefers lower values of the amplitude of linear perturbations \citep{Planck_2015_XIII}.

%% Previous work on coupled dark energy
If the growth of cosmological structure is confirmed to deviate from the theoretical predictions of the $\Lambda $CDM model, 
{one interpretation of this result would be  }%one interpretation of this result would be 
the discovery of a new regime of gravitational physics. However, {many of the most popular modified gravity theories, such as $f(R)$, Symmetron, and nDGP models, generically lead to a strengthening of the gravitational force. This naturally implies an enhanced growth of perturbations, in contrast with the observations which favour a suppressed growth rate. In this respect, it appears more}
plausible that a non-gravitational interaction is responsible for the anomalous behaviour.  Furthermore, in light of the stringent constraints on General Relativity derived from solar system measurements \citep[][]{Bertotti_Iess_Tortora_2003,Will_2005}, and further restrictions derived from the propagation of gravitational waves \citep{Lombriser_Lima_2016}, the latter options would also appear to offer a more natural solution. 

Many models of coupled dark energy have been proposed in the literature \citep[see e.g.][]{Amendola_2000,Barrow_Clifton_2006,Baldi_2011a}. However the overwhelming majority focus on a specific form of energy-momentum exchange between the two fluids, in which the coupling current is timelike. In other words, the theoretical models predominantly take the form of energy exchange rather than momentum exchange.  Motivated by the tendency for low-energy interactions between Standard Model particles to result in elastic scattering, \citet{Simpson_2010} proposed a model which invokes pure momentum exchange between the two fluids. Subsequently,  \cite{Pourtsidou_etal_2013} presented a  comprehensive classification of interacting Dark Energy models where they identified  a class of models (termed {\em `Type 3'}) which invoke  pure momentum exchange between dark matter and a scalar field. The properties of these models were explored in greater detail in \cite{Skordis_Pourtsidou_Copeland_2015}.   In this work we will aim to develop our understanding of the relationship between the Type 3 models and the elastic scattering model, and explore their phenomenological effects. 

%% Previous work on simulations of coupled dark energy
In \citet{Baldi_Simpson_2015} it was shown that if the dark matter particles experience a drag force as they pass through a canonical (i.e. non-phantom) dark energy fluid, they leave two key observational signatures. First of all the matter power spectrum is  suppressed  in a scale-independent fashion on linear scales, and secondly it is  scale-dependently enhanced on nonlinear scales. Here we present results from a suite of N-body simulations in which the dark energy fluid has an evolving equation of state, as is to be expected from dynamical models, and compare the resulting large scale structure with those found in \citet{Baldi_Simpson_2015}.  Our aim is then to test whether such a class of cosmologies might provide a way to reconcile cosmological constraints arising from CMB data analysis and low-redshift measurements of the growth of structures

%% Summary of sections
{In \S \ref{sec:scalar} we explore the relationship between the phenomenological dark scattering model, and those which invoke a velocity coupling to the derivative of a scalar field. Then in \S \ref{sec:models} we review the particular models chosen for further investigation using a suite of numerical simulations.  Specifications of the simulations are given in \S \ref{sec:simulations}, and their outputs are analysed in \S \ref{sec:results}. Our concluding remarks are presented in \S \ref{sec:concl}. 
}

\begin{figure}
\includegraphics[width=\columnwidth]{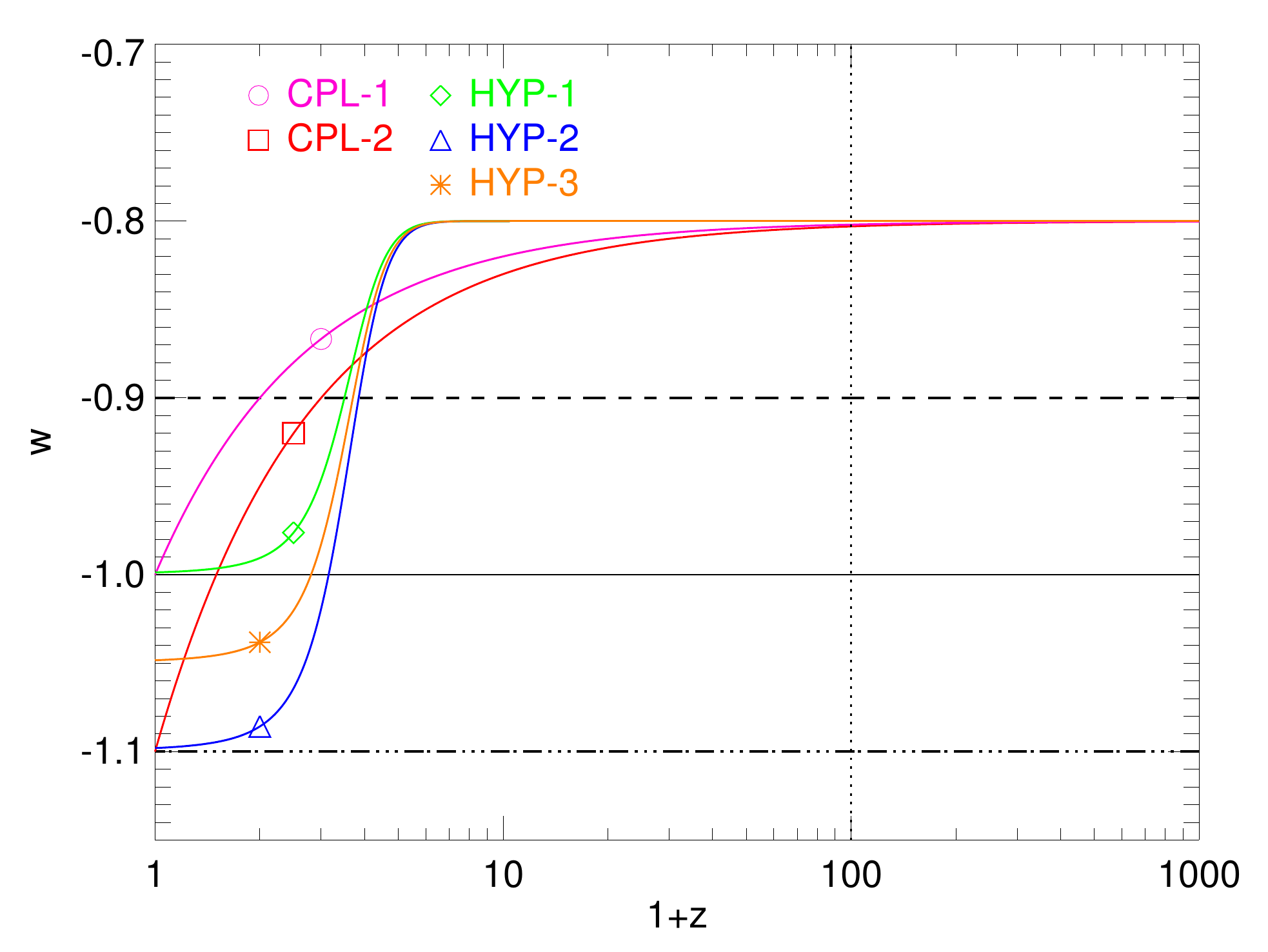}
\caption{{ The equation of state evolution for the various parameterisations considered in the present work. The black dashed and triple-dot-dashed lines represent the constant values $w=-0.9$ and $w=-1.1$, respectively, that have been discussed in \citet{Baldi_Simpson_2015}}}
\label{fig:eos}
\end{figure}

\begin{table}
\begin{center}
\begin{tabular}{cccc|cc}
\hline
Parameterisation & $w_{0}$  &  $w_{a}$  & $z_{t} $& $\xi $ & $\sigma _{8}$\\
\hline
\hline
$\Lambda $CDM & -1 & -- & -- & -- & 0.83 \\
\hline
{\bf w09} & -0.9 & -- & -- & $10, 50$ & $0.79\,, 0.75 $\\
{\bf w11} & -1.1 & -- & -- & $10, 50$ & $0.85\,, 0.87 $\\
\hline
{\color{magenta} {\bf CPL-1}} & -1 & 0.2 & -- & $50$ & $0.75$ \\
{\color{red} {\bf CPL-2}} & -1.1 & 0.3 & -- & $50$ & $0.79$\\
{\color{green} {\bf HYP-1}} & -1 & 0.2 & 2.5 & $50$ & $0.78$\\
{\color{blue} {\bf HYP-2}} & -1.1 & 0.3 & 2.5 & $50$ & $0.83$\\
{\color{orange} {\bf HYP-3}} & -1.05 & 0.25 & 2.5 & $50$ & $0.81$\\
\hline
\hline
\end{tabular}
\end{center}
\caption{The various DE parameterisations considered in the present work with their main parameters and the resulting value of $\sigma _{8}$ at $z=0$.}
\label{tab:models}
\end{table}

\section{Momentum Exchange with Scalar Fields}
\label{sec:scalar}

{
\citet{Pourtsidou_etal_2013} present three classifications of coupled dark energy models. Here we shall focus on the \emph{Type 3} class of models, which generate an exchange of momentum between dark matter and dark energy. This is achieved by invoking a coupling between the dark matter velocity field $u^\mu$ and the covariant derivative of a scalar field $\phi$, as follows

\[
S_\phi = - \int \ud^4 x \sqrt{-g} \left[F(Y) + F(Z)  + V(\phi)\right]  \, ,
\]
where the kinetic and velocity coupling terms are defined by
\[ \eqalign{
Y &\equiv \frac{1}{2} \nabla_\mu \phi  \nabla^\mu \phi   \, , \cr
Z &\equiv u^\mu \nabla_\mu \phi \, , \cr
}
\]

Note that in defining a continuous velocity field, as is required to form $Z$, we necessarily introduce a smoothing length over which the particle velocities are averaged.  The smoothing length is assumed to be smaller than the cosmological perturbations under consideration. 
%This leads to a ***

The momentum flux $\current$ is given by \cite{Skordis_Pourtsidou_Copeland_2015}:

 \[
\current = B_3 \delta_{\text{DE}}  + B_5 \theta_{\text{DE}} + B_6 \theta_c
\]
where $\delta$ and $\theta$ denote the density and velocity perturbations, and their three coefficients are  

\[ \label{eq:Bterms}
\eqalign{
B_3 &=  \frac{1}{1 -\frac{\zbar F_Z}{\rho_c }} \frac{\zbar F_Z c_s^2}{1+w} \, , \cr
B_5 &=  \frac{a}{1 -\frac{\zbar F_Z}{\rho_c }} \left[ \bar{X} \left(\frac{F_Z}{F_Y} - Z \right) + F_Z \left( \frac{\mu}{a F_Z} - \frac{F_\phi}{F_Y} \right) \right]      \, , \cr
B_6 & = -B_5 + \frac{3 \mathcal{H} \zbar F_Z c_s^2}{1 - \frac{\zbar F_Z}{\rho_c }}  \, .
}
\]
% \[ 
% \current = \frac{1}{1 - \frac{ \zbar F_Z}{\bar{\rho}_c}} \left[ \xbar  \psi - (\xbar \dot{\bar{\phi}} + F_Z \dot{\bar{Z}}) \theta_c 
 %   + F_Z  \delta Z \right]
% \]
Here subscripts denote derivatives, for example $F_Z \equiv d F(Z)/d Z$, and $c_s$ denotes the sound speed, while 
\[ \label{eq:X}
\eqalign{
\bar{X} &= \frac{1}{a} \left[ (Z F_{ZY} - F_{ZZ} ) \dot{\bar{Z}} - F_{Z\phi} \dot{\phi} - 3 \mathcal{H} F_Z \right] \, . \cr
% \delta Z &= \left[ \frac{\mu}{2 \gmma \zbar} - a F_\phi \right] \left[ \theta_{DE} + \frac{2 \gmma \zbar Z}{\rho_{DE}(1+w)} \theta_c \right] + \frac{c_s^2 \zbar}{1+w} \delta_{DE} \, \cr
 % \mu & = 6 \gmma \left(c_s^2 - c_a^2 \right) \left(1 + w \right) \bar{\rho}_{DE}   \mathcal{H}  \, .
}
\]

% Now taking the limit $\gmma \rightarrow 0.5$, this causes the effective sound speed to become vanishingly small,  $c_s^2 \rightarrow 0$.  This yields the following coefficients, from (86) of \citet{2015Skordis}:

In \citet{Skordis_Pourtsidou_Copeland_2015} the authors demonstrate that a formal equivalence cannot be drawn between the Type 3 models and the elastic scattering case. This is understandable  given that the velocity coupling $Z$ is associated with the gradient of the field, in contrast to the scattering model where the interaction is associated with  the local energy density. However,  for a particular subclass of Type 3 models, the characteristic drag-like behaviour can be reproduced. Provided the derivative of $F(Z)$ is large, such that $|F_Z| \gg |Z|$, as could occur for a variety of functions such as $F(Z) \propto \exp(-Z)$, then the expression simplifies considerably. There is only one contribution which has no explicit dependence on $Z$, which stems from the final term in equation (\ref{eq:X}), so to leading order we have

  \[
  \current =  3 \mathcal{H} F_Z^2 (\theta_c - \theta_{DE} ) + \mathcal{O}(Z)
  \]
%
% \[
% \mu = -3 \rho_{DE}(1+w)\mathcal{H} c_a^2 \, . 
% \]
By comparison, in the elastic scattering model  \citep{Simpson_2010} an expression is found which is also proportional to the difference in the velocity perturbations of the two fluids
\[
\current =  - \rho_{DE} (1 + w) a n_D \sigma_D  (\theta_c - \theta_{DE} )
\]
where $n_D \equiv n_0 a^{-3}$ is the proper number density of dark matter particles, $w\equiv p/\rho$ is the dark energy equation of state, and $ \sigma_D$ is the scattering cross-section. From the above, and utilising $\rho_{DE}(1 + w) \simeq  -Z F_Z $, we can define an effective cross-section as follows

% \[
% \sigma_{\eff} \equiv \frac{3 \mathcal{H} a^2 F_Z^2 }{\rho_{DE}  n_0 (1 + w)}  \, .
% \]
% We also note that $\rho_{DE}(1 + w) \simeq  -Z F_Z $, simplifying the expression  to 
\[
\sigma_{\eff} \equiv - \frac{3   \mathcal{H} a^2  F_Z }{   n_0 Z }  \, .
\]

% \[
% \sigma_{\eff} \equiv \frac{3 \gmma H a (c_a^2 - c_s^2)}{n_0}  \, .
% \]
% We generally expect $c_a^2 < c_s^2$, and thus we require  $\gmma < 0$ in order to mimic the phenomenology of the elastic scattering model. 

% Note that a negative value of $\gamma_0$ is also essential in order to ensure that the alignment of the rest frames of the two fluids represents a stable configuration - a minimum in the effective potential, as opposed to an unstable maximum. 

Furthermore, in the limit of weak coupling, we find that all Type 3 models generate a scale-independent modification to the linear growth of cosmic structure. This is due to the fact that, on scales smaller than the dark energy sound horizon, the dark energy  perturbations are driven by the potential well associated with the dark matter perturbations.  The density and velocity fields therefore all display approximately the same spatial distributions: $\delta_m(x) \propto \theta_m(x) \propto \delta_{\text{DE}}(x)  \propto \theta_{DE}(x)$ (see also eq 105 of \citet{Pourtsidou_etal_2013}). As a result, even in the most general form of Type 3 models $(S = B_3 \delta_{\text{DE}}  + B_5 \theta_{\text{DE}} + B_6 \theta_c)$ are well described on  sub-horizon scales $(k \gg k_H)$ by 
\[
S \propto \theta_m \, ,
\]
The microphysical interpretation is that the dark matter particles  will experience a force directly proportional to, and (anti-)parallel with, their velocity vector.  And it is this core phenomenological effect that we shall replicate within our numerical simulations. 
}

\section{Dark Scattering Models beyond a constant equation of state: two Dark Energy parameterisations}
\label{sec:models}

Motivated by the above outlined relation between models of elastic scattering in the dark sector and the particular sub-class of {\em `Type 3'} coupled quintessence models proposed in \citet{Skordis_Pourtsidou_Copeland_2015}, we
briefly review in this Section the main features of Dark Scattering cosmologies. {We also present} the specific models under investigation in the present work, providing an overview of their background evolution and of their main features related to linear and nonlinear structure formation.

\subsection{Background evolution}

We will consider cosmological models characterised by an exchange of momentum between Cold Dark Matter  particles {and} a Dark Energy field, modelled as a nearly-homogeneous fluid with a time-dependent equation of state parameter $w(a)$. In the present work we will consider two possible parameterisations of a {\em freezing} DE equation of state parameter $w(a)$. 

The first one is the standard and widely employed
Chevalier-Polarski-Linder parameterisation \citep[CPL hereafter,][]{Chevallier_Polarski_2001,Linder_2003}:
\begin{equation}
\label{CPL}
w_{\rm CPL}(a) \equiv w_{0} + (1-a)w_{a}\,,
\end{equation}
where $w_{0}$ is the value of the equation of state parameter at the present time and the parameter $w_{a}$ defines the
low-redshift evolution of $w_{\rm CPL}$. It should be noticed here that the CPL parameterisation provides an evolution of $w_{\rm CPL}$ ranging between $w_{0}$ at $z=0$ and $(w_{0}+w_{a})$ for $z\rightarrow \infty$, with a negative convexity.

However, as it has been recently pointed out in \citet{Pantazis_Nesseris_Perivolaropoulos_2016}, freezing models with a transition between a convex and concave shape might provide a less biased fit of observational data. Therefore,
the second parameterisation that we will consider in this work  will span the same global range of {our first model}%the CPL formula
, but be characterised by a shallower behaviour at very low redshifts and a sharper transition to the high-$z$ asymptotic value taking place at some intermediate redshift $z_{t}$ (that represents an additional free parameter) { which sets the point of inflection.}
%where the convexity of the $w(z)$ function changes sign
 Such behavior can be modelled by a hyperbolic tangent function of the form\footnote{A similar shape of the equation of state $w(z)$ has been recently proposed also by \citet{Jaber_Delamacorra_2016} using a different functional parameterisation.}:
\begin{equation}
\label{HYP}
w_{\rm HYP}(a) \equiv w_{0} + \frac{w_{a}}{2}\tanh \left(\frac{1}{a} - z_{t}\right) .
\end{equation}

The evolution of the equation of state parameter as a function of redshift for these two parameterisations is shown in Fig.~\ref{fig:eos} for the parameters summarised in Table~\ref{tab:models}. In Fig.~\ref{fig:eos}, like in all the figures of the present work, we also display for comparison the behavior of the two constant-$w$ models with $w=-0.9$ and $w=-1.1$ that were studied in our previous paper \citet{Baldi_Simpson_2015} (as dashed and triple-dot-dashed lines, respectively).

\begin{figure}
\includegraphics[width=\columnwidth]{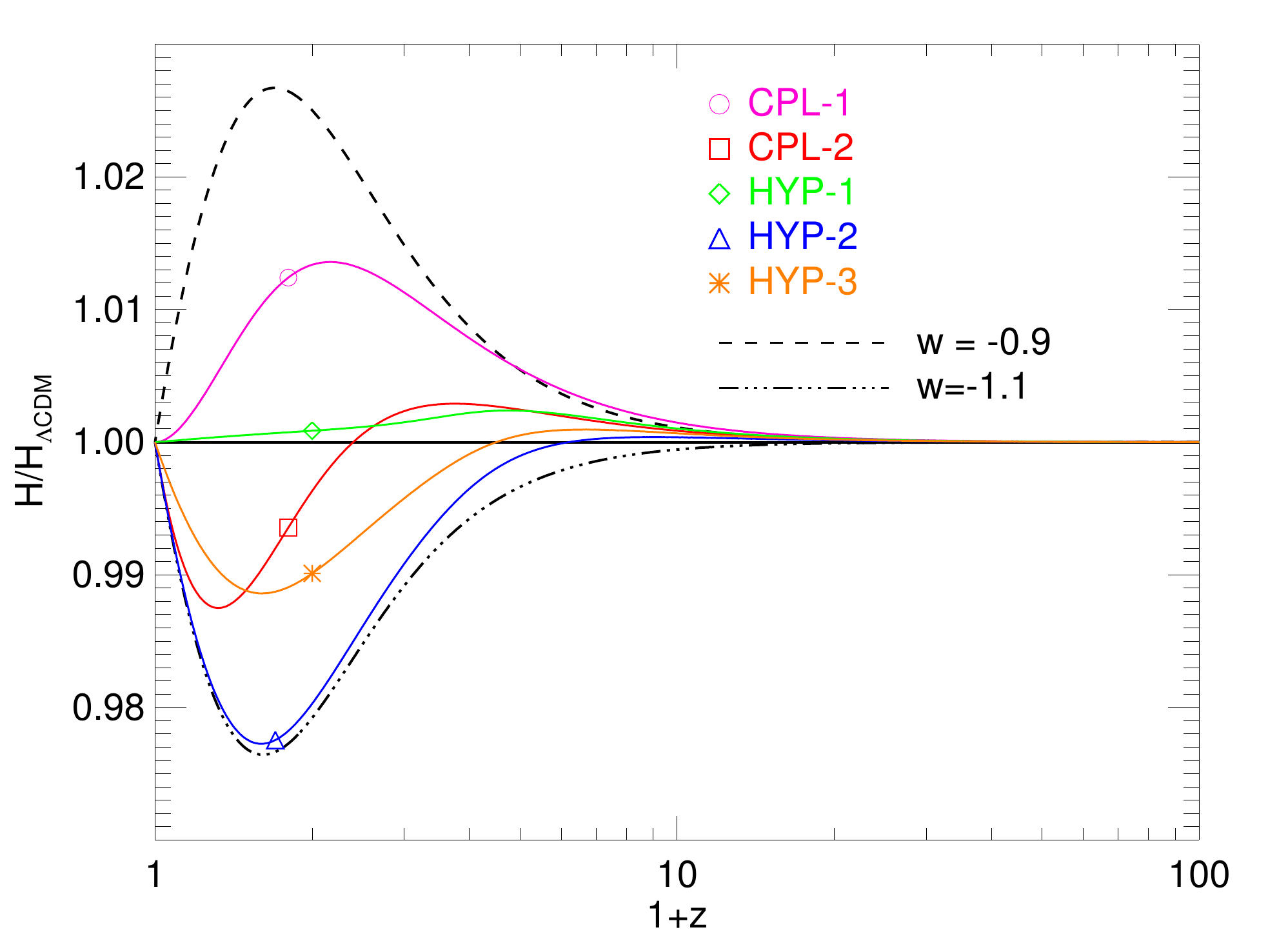}
\caption{{ The Hubble function ratio to the standard $\Lambda $CDM case for the various parameterisations considered in the present work. The black solid and dashed lines represent the constant values $w=-0.9$ and $w=-1.1$, respectively, that have been discussed in \citet{Baldi_Simpson_2015}}}
\label{fig:hubble}
\end{figure}

The background expansion of the various models will be described by the Hubble function $H(z)$:
\begin{align}
H^{2}(z) =& H_{0}\left[ \Omega _{M}(1+z)^{2} + \Omega _{r}(1+z)^{4} + \Omega _{K}(1+z)^{2} +  \right.\nonumber \\
& \left. \Omega _{\rm DE} e^{\int_{0}^{z}\frac{3(1+w(\tilde{z}))}{1+\tilde{z}}d\tilde{z}}\right] 
\label{hubble}
\end{align}
which is displayed in Fig.~\ref{fig:hubble}, along with its ratio to the standard $\Lambda $CDM case.
In Eq.~\ref{hubble}, the dimensionless density parameters $\Omega _{i}\equiv \rho _{i}/\rho _{crit}$ refer to the components of matter ($M$), radition ($r$), curvature ($K$) and dark energy (DE), with the critical density of the universe being $\rho _{crit} \equiv 3H^{2}/8\pi G$. As one can see from the figure, all the models considered in this work do not deviate by more than 2.5\% from the $\Lambda $CDM expansion history, with a maximum deviation around $z\approx 1$, and the variable-$w$ models are found to be all closer to $\Lambda $CDM compared to the two constant-$w$ scenarios investigated in \citet{Baldi_Simpson_2015}. Therefore, these models very closely resemble the standard cosmological scenario at the level of the background evolution.

\begin{figure}
\includegraphics[width=\columnwidth]{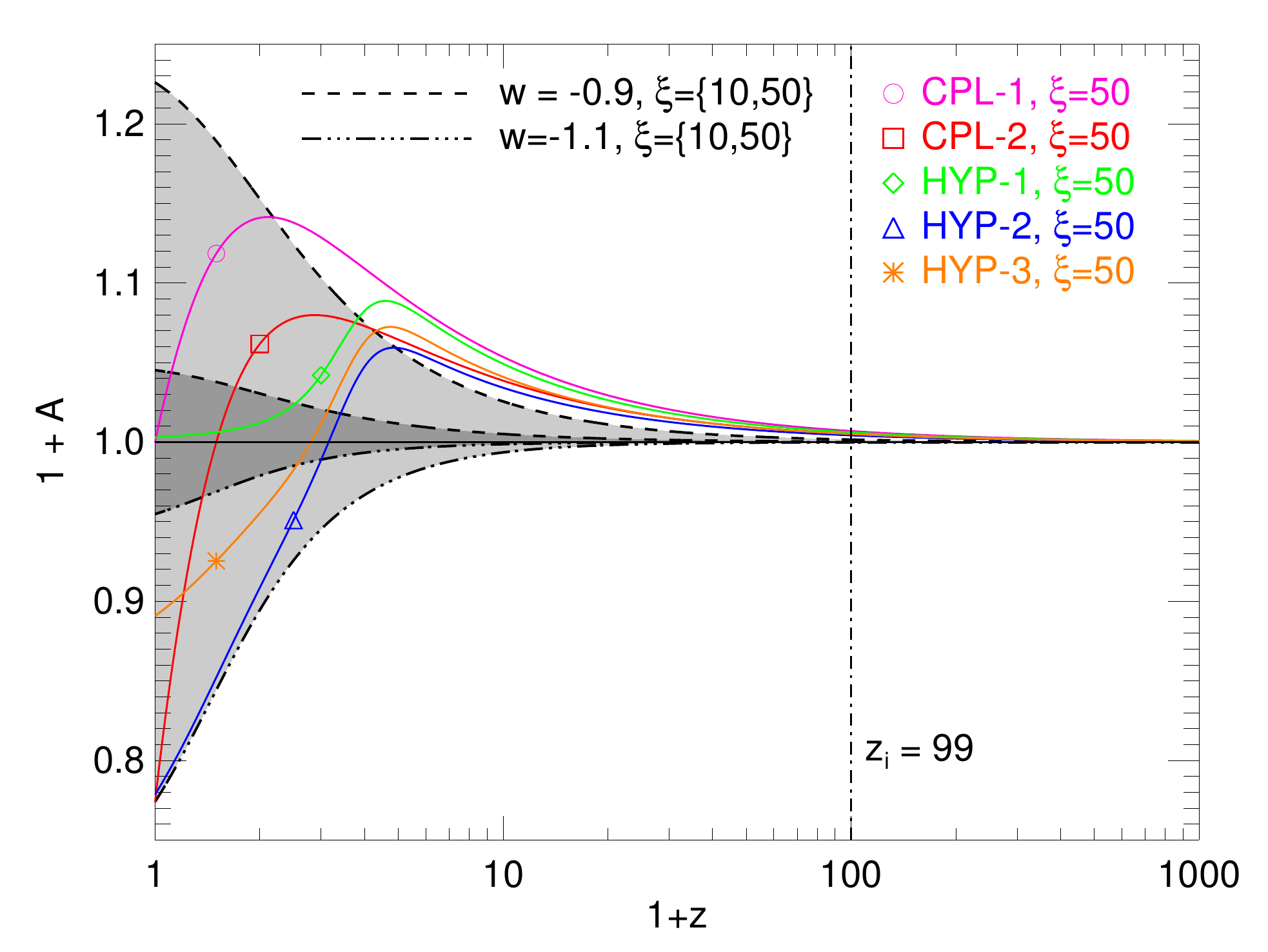}
\caption{{ The modified friction term $(1+A)$ for the various parameterisations considered in the present work. As on can see in the figure, the variable-$w$ models determine a stronger modification at high redshifts and a weaker modification at low redshifts compared with the constant-$w$ case. Colors and linestyles are the same as in Fig.~\ref{fig:hubble}}}
\label{fig:friction_factor}
\end{figure}
 
\subsection{Linear and non-linear structure formation in Dark Scattering cosmologies}

As already extensively discussed in \citet{Simpson_2010} and \citet{Baldi_Simpson_2015}, the evolution of linear perturbations in the presence of a momentum exchange between dark energy and CDM particles is described by the system of coupled equations in Fourier space\footnote{In the present paper -- if not stated otherwise -- we will always assume units in which the speed of light is unity, $c=1$.}:
\begin{equation}
\label{eq:velocity_Q}
\eqalign{
\theta'_{\rm DE} &= 2 H \theta_{\rm DE} - a n_{\rm CDM} \sigma _{\rm D} \Delta \theta + k^2 \phi + k^2 \frac{\delta_{\rm DE}}{1+w}   \, , \cr
\theta'_{\rm CDM}  &= - H \theta_{\rm CDM} + \frac{\rho_{\rm DE}}{\rho_{\rm CDM}} (1+w) a n_{\rm CDM} \sigma _{\rm D} \Delta \theta + k^2 \phi  \, ,
}
\end{equation}
\noindent where $n_{\rm CDM}$ is the proper number density of CDM particles, $\Delta \theta \equiv \theta_{\rm DE} - \theta_{\rm CDM}$ is the velocity contrast,  ($\theta _{i}$ being the divergence of the velocity perturbations for the field $i$), $\phi $ is the gravitational potential sourced by the Poisson equation $k^{2}\phi = 4\pi G (\delta _{\rm CDM} + \delta _{\rm DE})$, and a prime denotes a derivative with respect to cosmic time. 

By assuming a DE sound speed $c_{s}^{2} = 1$, which is predicted by most DE models based on light scalar fields, we can expect DE perturbations to be damped within the cosmic horizon so that the DE density and velocity fields are approximately homogeneous (i.e. $\delta _{\rm DE}=\theta _{\rm DE}=0$), {as was confirmed numerically in the previous paper}. Therefore, as we will concentrate on sub-horizon scales, we shall neglect the influence of dark energy perturbations within the simulation and safely approximate $\Delta \theta \approx -\theta _{\rm CDM}$. With such approximation the linear Euler equation for CDM becomes:
\begin{equation}
\theta'_{\rm CDM}  = - H \theta _{\rm CDM} \left[ 1 + A \right] + k^2 \phi  \, ,
\end{equation}
where the second term in the brackets is the additional friction associated with the momentum exchange, defined as:
\begin{align}
\label{drag_1}
A\equiv &\frac{\rho_{\rm DE}}{H\rho _{\rm CDM}} (1+w) n_{\rm CDM} \sigma_{\rm D} = \nonumber \\
&\left( 1 + w\right) \frac{\sigma_{\rm D}}{m_{\rm CDM}} \frac{3 \Omega _{\rm DE}}{8\pi G} H \,.
\end{align}
This extra drag force depends on three free quantities: the DE equation of state $w(z)$, the DE-CDM scattering cross section $\sigma _{\rm D}$, and the CDM particle mass $m_{\rm CDM}$. In particular, the overall magnitude of the drag force depends on the latter two parameters only through their ratio, so that we can define the combined quantity 
{
\begin{equation}
\label{xi}
%\xi \equiv \frac{c\cdot \sigma _{\rm D}}{m_{\rm CDM}}
\xi \equiv \frac{\sigma _{\rm D}}{m_{\rm CDM}}
\end{equation}}
with dimensions of 
%$[{\rm bn} \cdot {\rm c}^{3}/{\rm GeV}]$, 
$[{\rm bn}/{\rm GeV}]$,
as the main characteristic parameter of our models, such that Eq.~\ref{drag_1} becomes:
\begin{equation}
\label{drag_2}
A\equiv \left( 1 + w\right) \frac{3 \Omega _{\rm DE}}{8\pi G} H \xi \,.
\end{equation}
From Eqs.~\ref{drag_1},\ref{drag_2}, we notice that the additional term $A$ can be both positive or negative (i.e. acting as a friction or as a dragging force) for values of the DE equation of state $w$ above or below the cosmological constant value $w=-1$, respectively. While in \citet{Baldi_Simpson_2015} we focused on the simplified case of a constant equation of state, investigating the two cases $w=\left\{ -0.9,-1.1\right\}$ for different values of the parameter $\xi $, in the present work we aim to go beyond such rather unrealistic assumption and test Dark Scattering scenarios with variable $w(z)$, focusing on the few models described in Table~\ref{tab:models}, while keeping fixed the value of $\xi $.\\
%$\xi = 50 [{\rm bn} \cdot {\rm c}^{3}/{\rm GeV}]$.\\

A new friction term, analogous to the Thomson drag force experienced by electrons, is introduced to the equation of motion for individual CDM particles. 
\begin{equation}
\label{acceleration}
\dot{\vec{v}}_{i} = -\left[ 1 + A \right] H \vec{v}_{i} + \sum_{j \ne i}\frac{Gm_{j}\vec{r}_{ij}}{|\vec{r}_{ij}|^{3}}
\end{equation}
where $\vec{r}_{ij}$ is the distance between the $i$-th and the $j$-th particle.
The evolution of the factor $(1+A)$ is shown in Fig.~\ref{fig:friction_factor} for the different variable-$w$ models under investigation with a value of $\xi = 50\, [{\rm bn}/{\rm GeV}]$, while the dashed and dot-dashed lines enclosing the dark-grey and light-grey shaded areas correspond to the case of a constant $w$ with $\xi = 10$ and $50\, [{\rm bn}/{\rm GeV}]$, respectively. As one can see by comparing Figs.~\ref{fig:hubble} and \ref{fig:friction_factor}, the DE parameterisations considered in this work have a weaker impact on the background expansion history and on low-redshift structure formation as compared to the constant-$w$ models investigated in \citet{Baldi_Simpson_2015}, while they are expected to have a stronger effect on the growth of structures at high redshifts. As we will see later in the paper, this will imply an overall weaker impact on most cosmological observables while retaining interesting and non-trivial effects on the abundance of massive clusters and on the expected weak lensing signal.

\begin{table}
\begin{center}
\begin{tabular}{cc}
\hline
Parameter & Value\\
\hline
$H_{0}$ & 67.8 km s$^{-1}$ Mpc$^{-1}$\\
$\Omega _{\rm M} $ & 0.308 \\
$\Omega _{\rm DE} $ & 0.692 \\
$ \Omega _{b} $ &0.0482 \\
\hline
${\cal A}_{s}$ & $2.215 \times 10^{-9}$\\
$n_{s}$ & 0.966\\
\hline
\end{tabular}
\end{center}
\caption{A summary of the cosmological parameters adopted for all the simulations discussed in the present work.}
\label{tab:parameters}
\end{table}

\begin{figure*}
\includegraphics[scale=0.3]{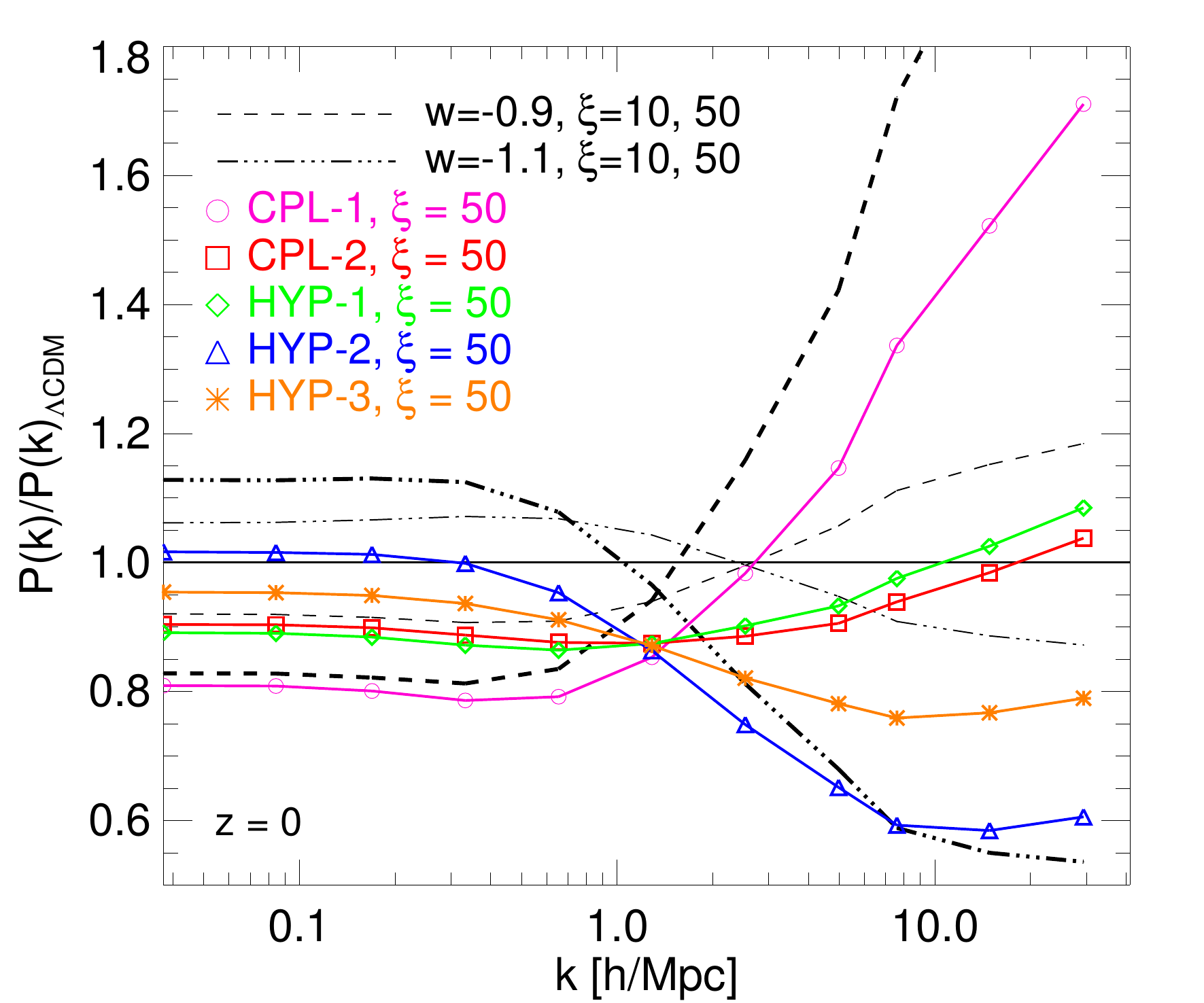}
\includegraphics[scale=0.3]{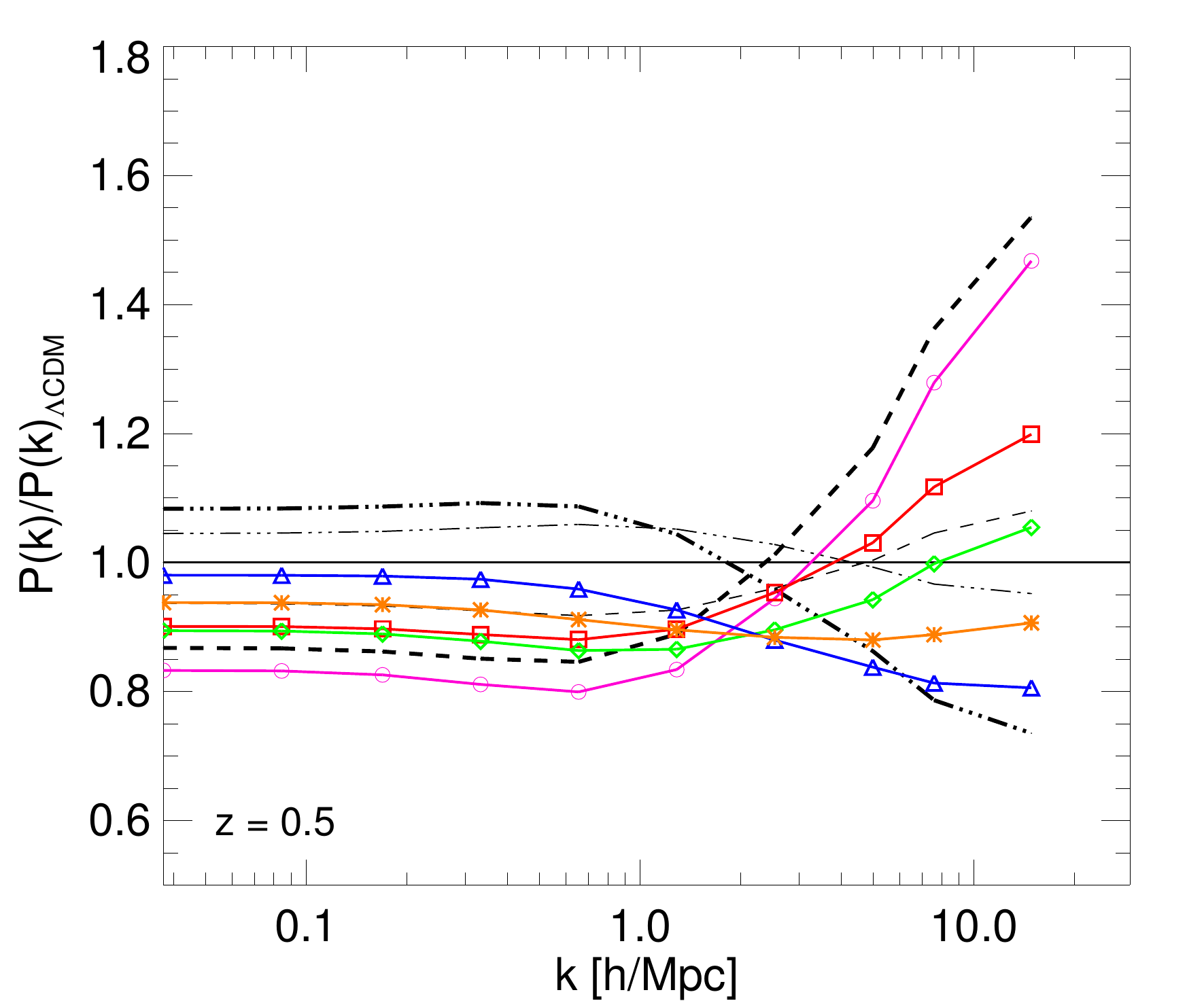}
\includegraphics[scale=0.3]{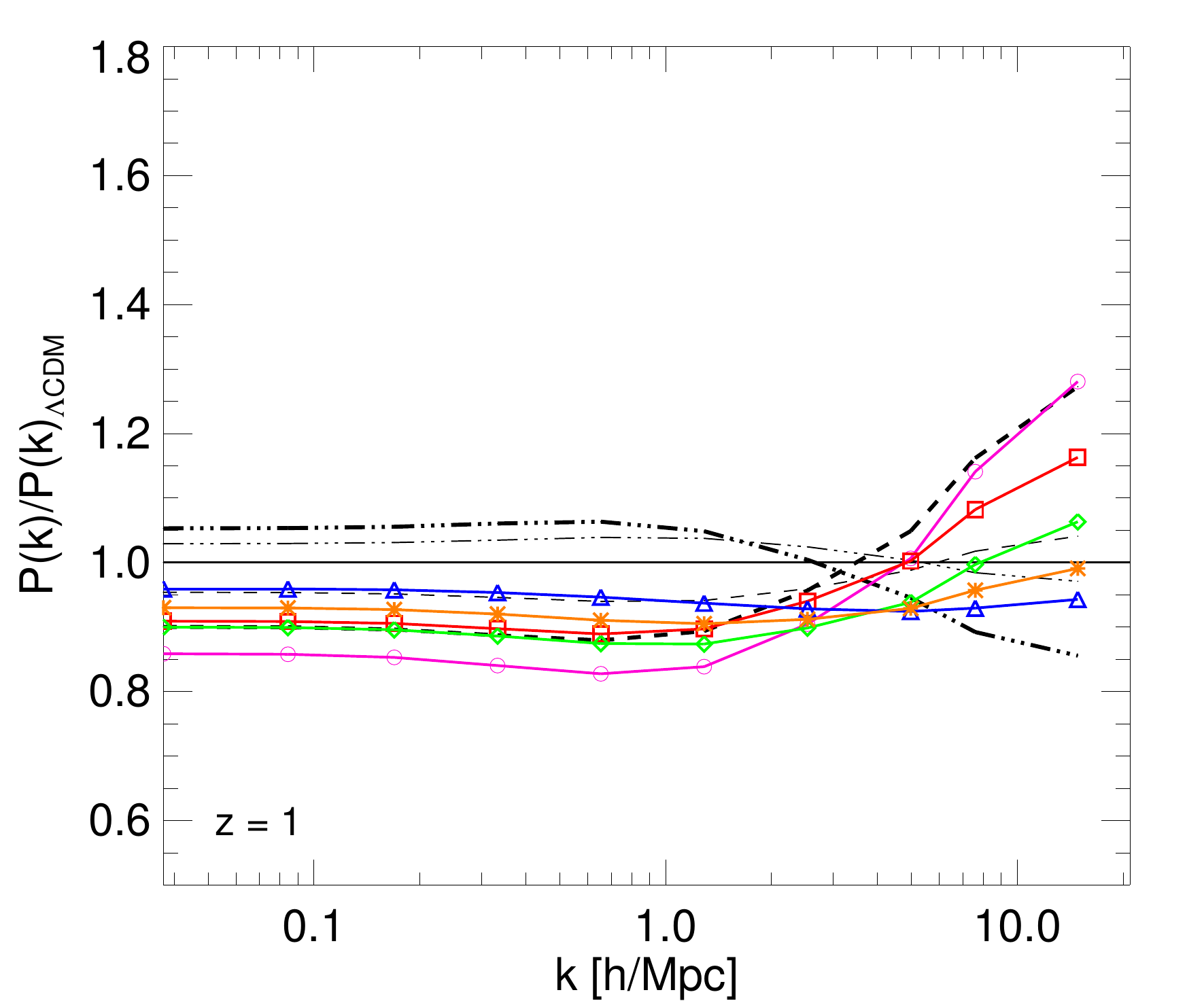}
\caption{{ The nonlinear matter power spectrum ratio to the reference $\Lambda$CDM model at three different redshifts $z=0$ ({\em left}), $z=0.5$ ({\em middle}), and $z=1$ ({\em right}) for the various models investigated with our intermediate-size simulations. The colours and linestyles are the same as in Fig.~\ref{fig:eos}}}
\label{fig:small_power_ratio_LCDM}
\end{figure*}
\begin{figure}
\includegraphics[width=\columnwidth]{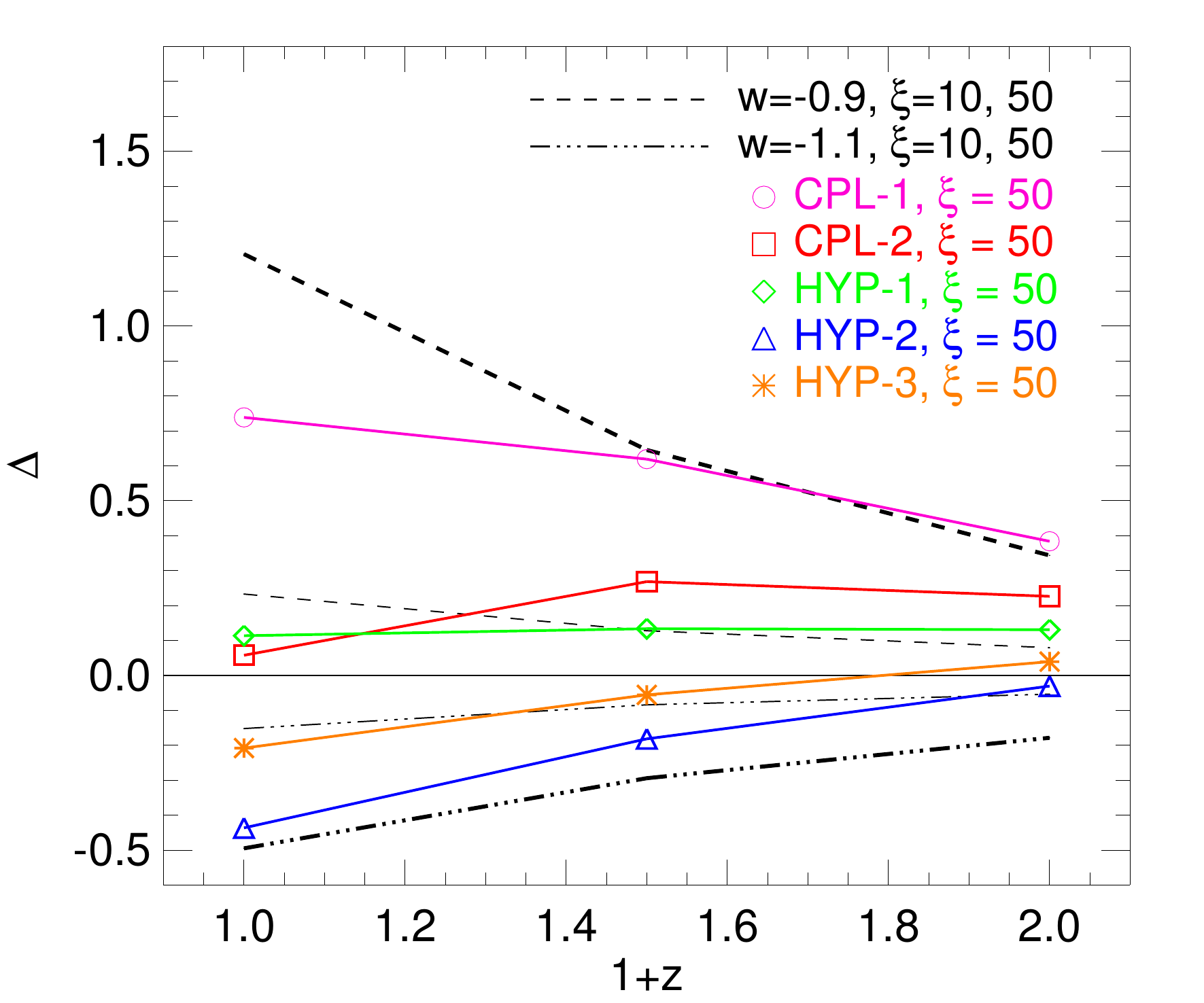}
\caption{ The relative difference between the nonlinear (i.e. measured at $k=10\, h/$Mpc) and the linear (i.e. measured at $k=0.1\, h/$Mpc) effects on the matter power spectrum for the various models of momentum exchange.}
\label{fig:lin-nonlin}
\end{figure}

\section{The Simulations}
\label{sec:simulations}

For all the models summarised in Table~\ref{tab:models}, and for a reference $\Lambda $CDM cosmology, we have performed a set of intermediate-size simulations with the modified version of {\small GADGET} \citep[][]{gadget-2} described in \citet{Baldi_Simpson_2015}, which self-consistently implements the drag force associated with the DE-CDM momentum exchange. These simulations have a box size of $250$ Mpc$/h$ a side and follow the evolution of $512^3$ CDM particles in a periodic cosmological volume from $z_{i}=99$ down to $z=0$. The mass resolution is $m = 1\times 10^{10}$ M$_{\odot }/h$ and the gravitational softening is $\epsilon \approx 12$ kpc$/h$. All the simulations share the same initial conditions (since the effect of the momentum exchange is negligible at $z\gtrsim 100$, see Fig.~\ref{fig:friction_factor}) and cosmological parameters \citep[consistent with the results of the Planck satellite mission,][see Table~\ref{tab:parameters}]{Planck_2015_XIII}.

This suite of simulations has been employed to test the effects of the variable-$w$ models on two basic cosmological observables: the nonlinear matter power spectrum and the halo mass function (see \ref{sec:medium}). These preliminary results allowed us to identify the most relevant sets of parameters for both the CPL and the HYP parameterisations to be investigated more extensively with a set of larger simulations. The latter are cosmological boxes of $1$ Gpc$/h$ a side filled with $1024^3$ CDM particles, and have therefore a poorer mass ($m = 8\times 10^{10}$ M$_{\odot }/h$) and space ($\epsilon \approx 24$ kpc$/h$) resolution compared to the smaller runs, but significantly improve the statistics of massive clusters and of cosmic voids (see \ref{sec:large}), thereby allowing a more significant assessment of the impact of the momentum exchange on the statistical and structural properties of these classes of objects.

For all the simulations initial conditions have been generated by displacing particle positions from a {\em ``glass"} homogeneous distribution \citep[][]{Davis_etal_1985} according to the Zeldovich approximation \citep[][]{Zeldovich_1970} to set up a random-phase realization
of the power spectrum predicted by {\small CAMB}\footnote{www.camb.info} \citep[][]{camb} for a $\Lambda $CDM cosmology with the chosen cosmological parameters. Therefore, all the differences that will be identified among the simulations outputs at low redshifts can be unambiguously ascribed to the effects of the different cosmological models. Furthermore, as the initial conditions are identical the comparison of the various models will not be affected by sample variance, and statistical uncertainties will be only due to Poisson noise.

\begin{figure*}
\includegraphics[scale=0.3]{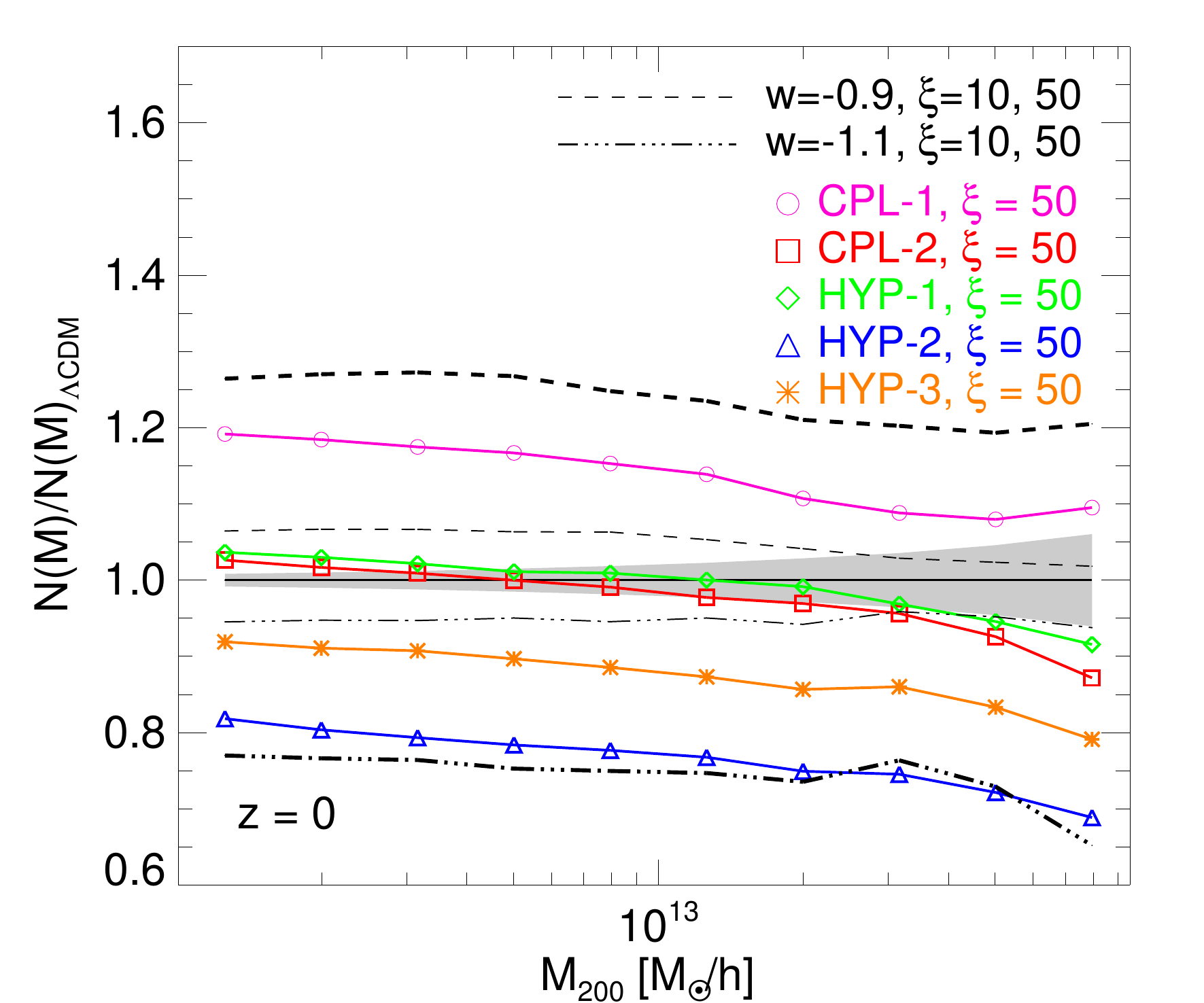}
\includegraphics[scale=0.3]{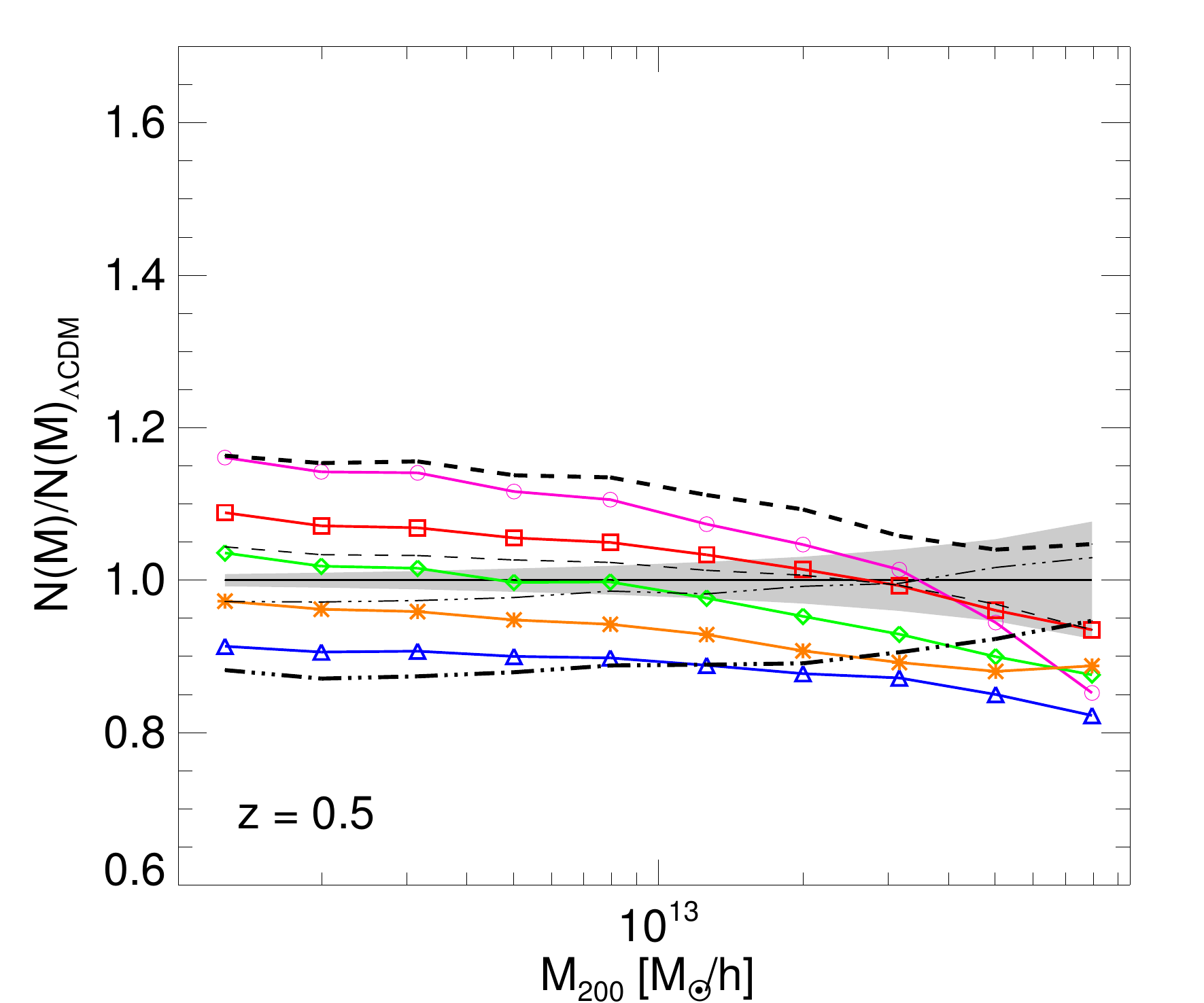}
\includegraphics[scale=0.3]{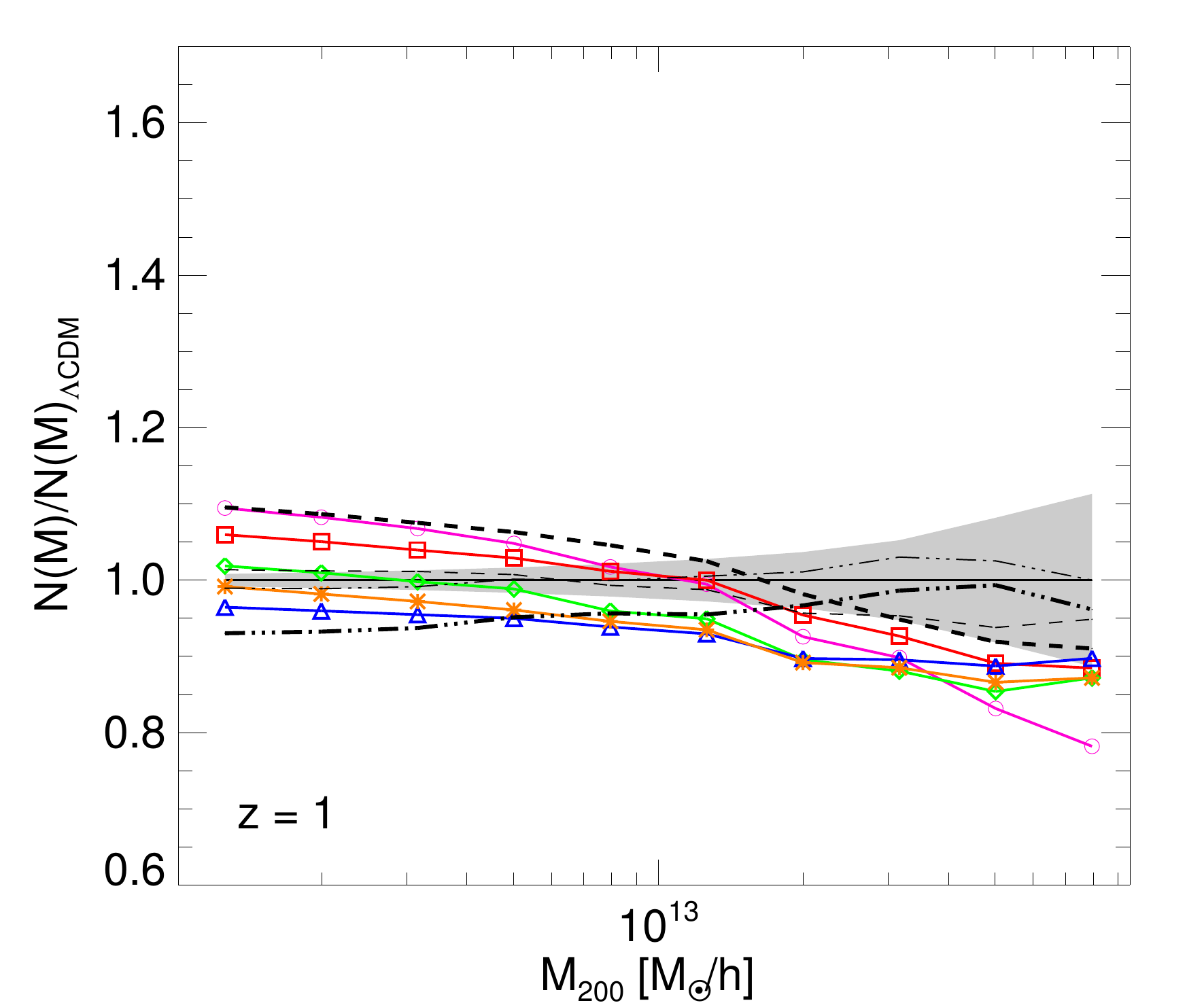}
\caption{{ The mass function ratio to the $\Lambda $CDM case for all the models under investigation in the present work. The three panels refer to the same redshifts considered above ($z=0\,, 0.5$ and $1$), and colours, symbols, and line styles are the same as displayed in Fig.~\ref{fig:small_power_ratio_LCDM}.}}
\label{fig:small_HMF_ratio_LCDM}
\end{figure*}

\begin{figure}
\begin{center}
\includegraphics[height=0.3\textheight]{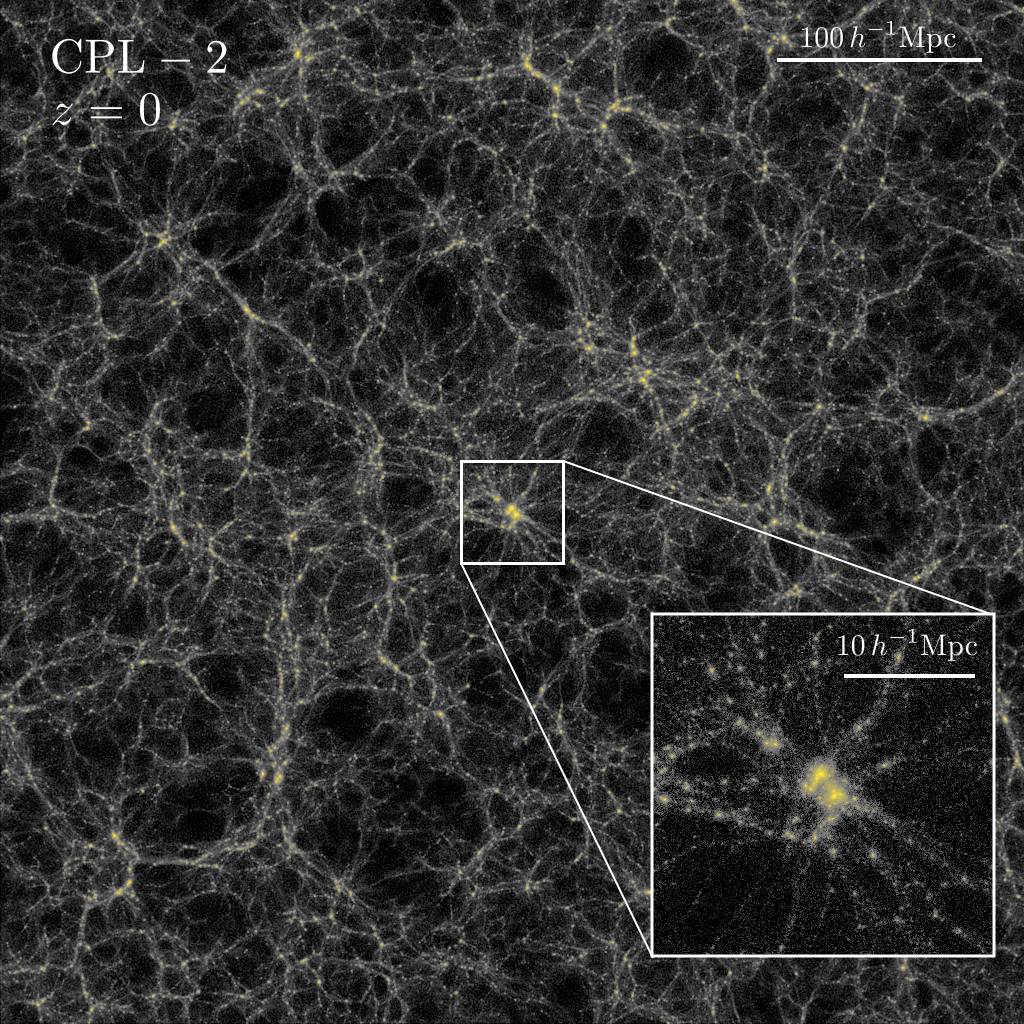}\\
\includegraphics[height=0.3\textheight]{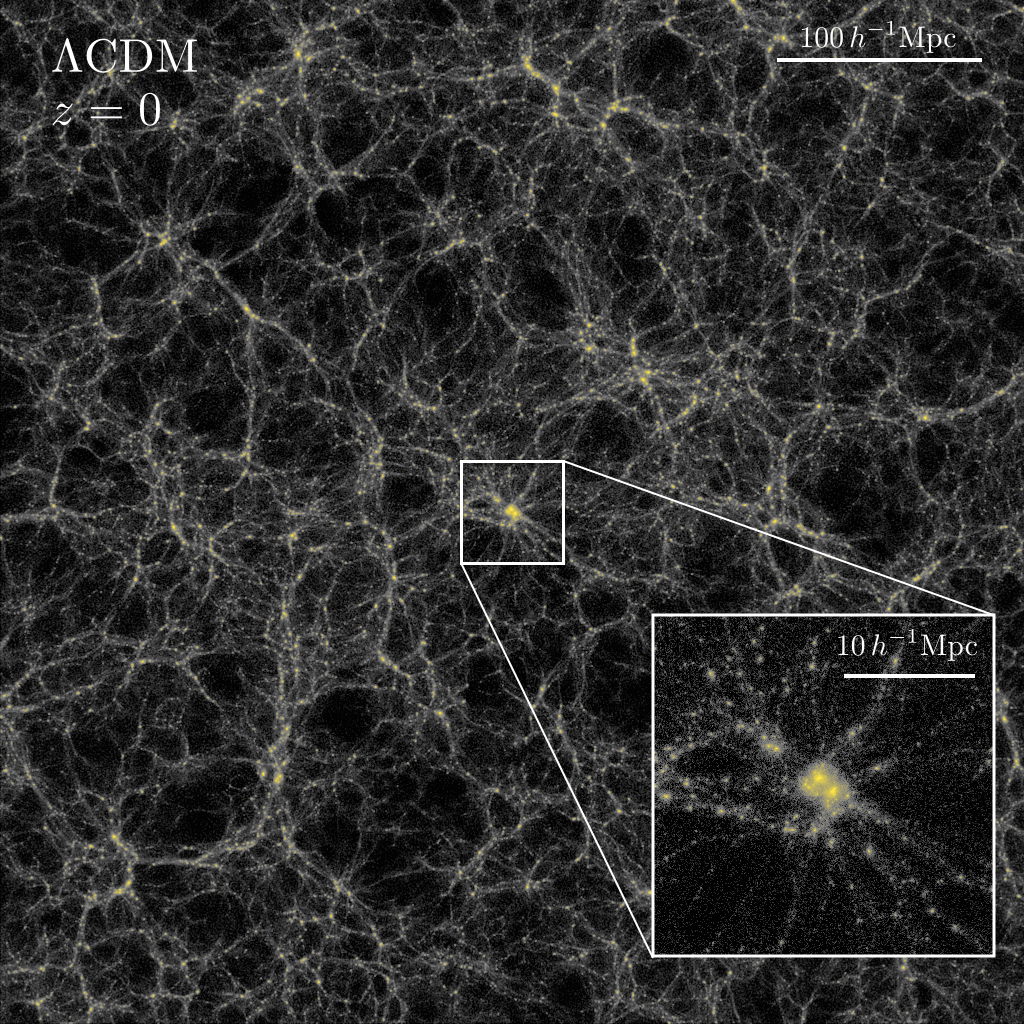}\\
\includegraphics[height=0.3\textheight]{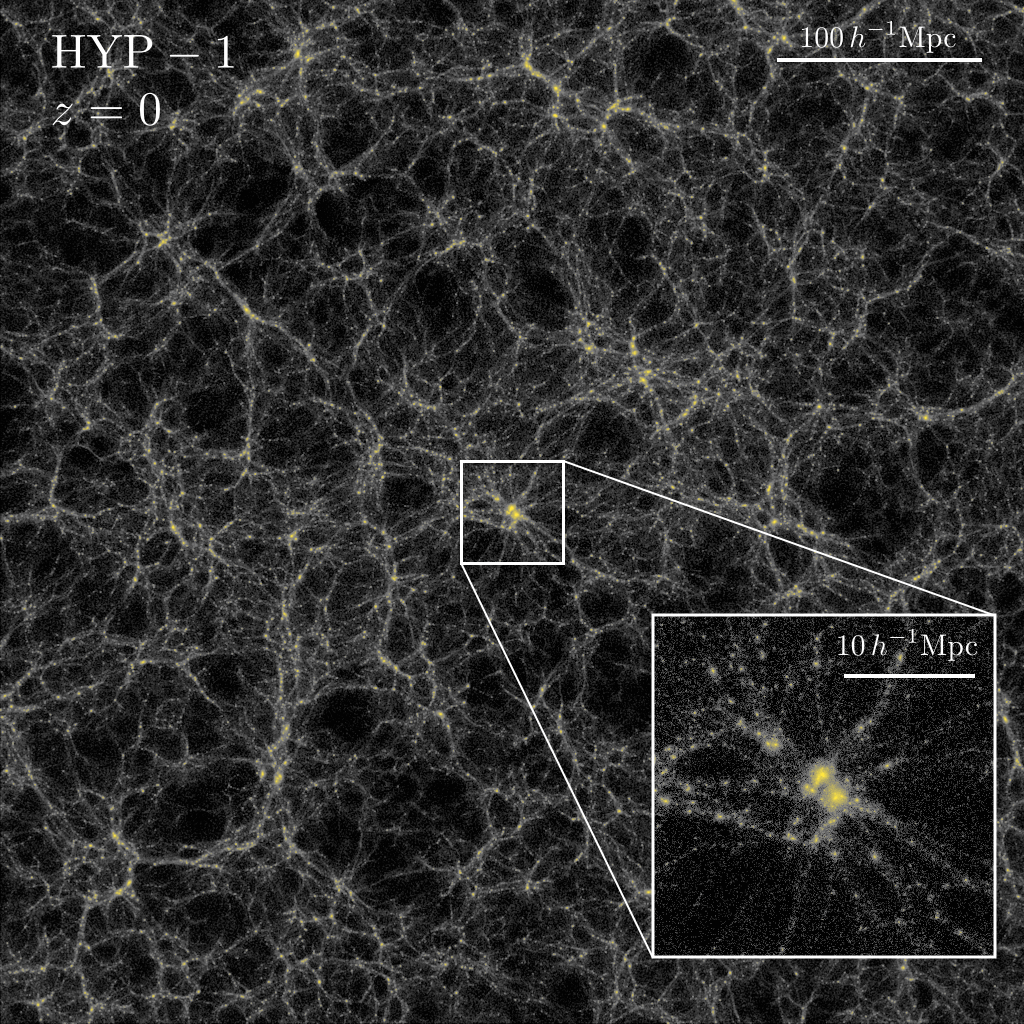}
\end{center}
\caption{{ The density field at $z=0$ in a slice of size $500$ Mpc$/h$ and thinkness $30$ Mpc$/h$ for the reference $\Lambda $CDM simulation (central panel) and for the selected CPL and HYP parameterisations (top and bottom panels, respectively). The slices have been centered on the most massive structure identified in the simulations which is displayed in the zoomed inset.}}
\label{fig:density_field}
\end{figure}

\section{Results}
\label{sec:results}

In the present section we will discuss the main outcomes of our simulations, starting with the results of the intermediate-scale runs and then moving to the large-scale realisations for the selected subset of models.

\subsection{Intermediate-size simulations: selecting target models through the nonlinear matter power spectrum and the halo mass function}
\label{sec:medium}

For all our intermediate-size simulations we extract the nonlinear matter power spectrum through a Cloud-in-Cell mass assignment to a cubic cartesian grid having the same spacing of the mesh used for the large-scale N-body integration, i.e. $512^{3}$ grid nodes. This allows to measure the power spectrum from the fundamental mode $k_{0} \approx 0.01$ $h/$Mpc up to the Nyquist frequency of the grid $k_{\rm Ny} \approx 6.43$ $h$/Mpc. In order to extend this range to smaller scales we employ the folding method of \citet{Jenkins_etal_1998,Colombi_etal_2009} and we smoothly interpolate the two estimations around $k_{\rm Ny}$. Then, the combined power spectrum obtained in this way is truncated at the scale where the shot noise reaches 20\% of the measured power. We apply this procedure to the simulation snapshots corresponding to three different redshifts $z=\left\{ 0\,, 0.5\,, 1\right\}$. The results are displayed in Fig.~\ref{fig:small_power_ratio_LCDM} where we show the ratio of the measured power of each simulation to the corresponding $\Lambda $CDM result. All the variable-$w$ cosmologies (solid coloured lines with symbols) are characterised by a value of $\xi = 50\, [{\rm bn}/{\rm GeV}]$ while for the two constant-$w$ models used as a reference (dashed and dot-dashed black lines) we consider both $\xi = 10\, [{\rm bn}/{\rm GeV}]$ (thin lines) and $\xi = 50\, [{\rm bn}/{\rm GeV}]$ (thick lines).

As one can see in the plots, and most evidently in the $z=0$ panel, the variability of the DE equation of state introduces non-trivial features in the behavior of the matter power as a function of scale. For constant-$w$ \citep[as already discussed in][]{Baldi_Simpson_2015} the effect of the DE-CDM momentum exchange on the power spectrum shows a scale-independent suppression (enhancement) of power at large scales and a transition to a scale-dependent enhancement (suppression) at small scales for $w>-1$ ($w<-1$). The transition corresponds to the scale where nonlinear effects come into play, and expectedly shifts towards smaller $k$ for decreasing redshift. Also, there is a direct correspondence between the magnitude of the linear and nonlinear effects, with a larger effect at linear scales being always associated with a larger effect also at nonlinear scales. 

For the variable-$w$ case the effects appear more diverse, with a wide range of behaviours and of possible linear-nonlinear transitions, as well as a less direct correspondence between the magnitude of the effect at linear scales and its nonlinear counterpart. 
In this respect, it is interesting to consider the relation between the power spectrum ratio at large scales and that  at smaller scales as the former will have mostly an impact on the statistical properties of large-scale structures while the latter will have direct consequences on the structural properties of collapsed objects. This comparison is shown in Fig.~\ref{fig:lin-nonlin}, where we display the relative difference of the observed power enhancement at $k=0.1\, h/$Mpc and at $k=10\, h/$Mpc defined as $\Delta \equiv [P(k)/P(k)_{\Lambda {\rm CDM}}]_{k=10}/[P(k)/P(k)_{\Lambda {\rm CDM}}]_{k=0.1} -1$, as a function of redshift. As one can see in the plot, all the variable-$w$ models have a weaker impact at nonlinear scales compared to the constant-$w$ models with the same value of the $\xi $ parameter ($\xi = 50\,  [{\rm bn}/{\rm GeV}]$, thick lines), and two of them (the CPL-2 and the HYP-1 models) even show a smaller nonlinear impact at $z=0$ compared to the constant-$w$ models with the lower value of $\xi = 10\,  [{\rm bn}/{\rm GeV}]$ (thin lines). This provides us with a useful guideline to select relevant combinations of parameters, since we are interested in identifying models that produce a sizeable suppression of the large-scale structures growth without  changing too dramatically the structural properties of collapsed halos.\\

A similar analysis can be performed using the abundance of halos as a test observable. To this end, we have identified particle groups in our simulations by means of a Friends-of-Friends algorithm with linking length $0.2$ times the mean inter-particle separation, and subsequently performed a particle unbinding on all the identified groups by means of the {\small SUBFIND} algorithm \citep[][]{Springel_etal_2001} in order to select gravitationally bound substructures. With these catalogs at hand, we have computed the cumulative halo mass function for all the models by binning the halos in mass bins according to their $M_{200}$ mass defined as the mass contained in a sphere centered on the most bound particle of each main substructure with a radius $R_{200}$ enclosing a mean density $200$ times larger than the critical density of the universe. We have then compared these mass functions to the outcome of the $\Lambda $CDM simulation, and the results are displayed in Fig.~\ref{fig:small_HMF_ratio_LCDM}.

As one can see in the figure, also in this case we find that the variable-$w$ models predict a milder impact on the abundance of halos at all masses compared to their constant-$w$ counterparts with the same value of $\xi = 50\,  [{\rm bn}/{\rm GeV}]$, especially at $z=0$. Furthermore, consistently with the previous findings shown in Figs.~\ref{fig:lin-nonlin} and \ref{fig:small_power_ratio_LCDM}, the two models CPL-2 and HYP-1 appear to be the closest match to the $\Lambda $CDM result at low masses, while showing some significant suppression of the abundance of halos at the high-mass end of the available catalogs, differently from all the other models that are found to determine large deviations in the abundance also for galaxy-sized halos. This is a consequence of the suppression of large-scale linear clustering, since the halo mass function is exponentially sensitive to the value of $\sigma _{8}$, and to the relatively weak impact of the interaction at highly nonlinear scales. Therefore, these models represent  promising % the best 
candidates to ease the persisting tensions between both the observed weak lensing amplitude \citep[][]{syspaper} and abundance of Planck SZ clusters \citep[][]{Planck_XXIV} on one side and their predicted values based on the maximum likelihood Planck 2015 cosmological parameters estimation \citep[][]{Planck_2015_XIII}. \\

Based on these preliminary checks, we have then selected the two models CPL-2 and HYP-1 to be investigated in more detail through larger simulations. along with two constant-$w$ models with $\xi = 10\,  [{\rm bn}/{\rm GeV}]$ and a reference $\Lambda $CDM cosmology. The outcomes of these larger simulations, that represent the core results of the present work, are discussed in the following Section.

\begin{figure*}
\includegraphics[scale=0.3]{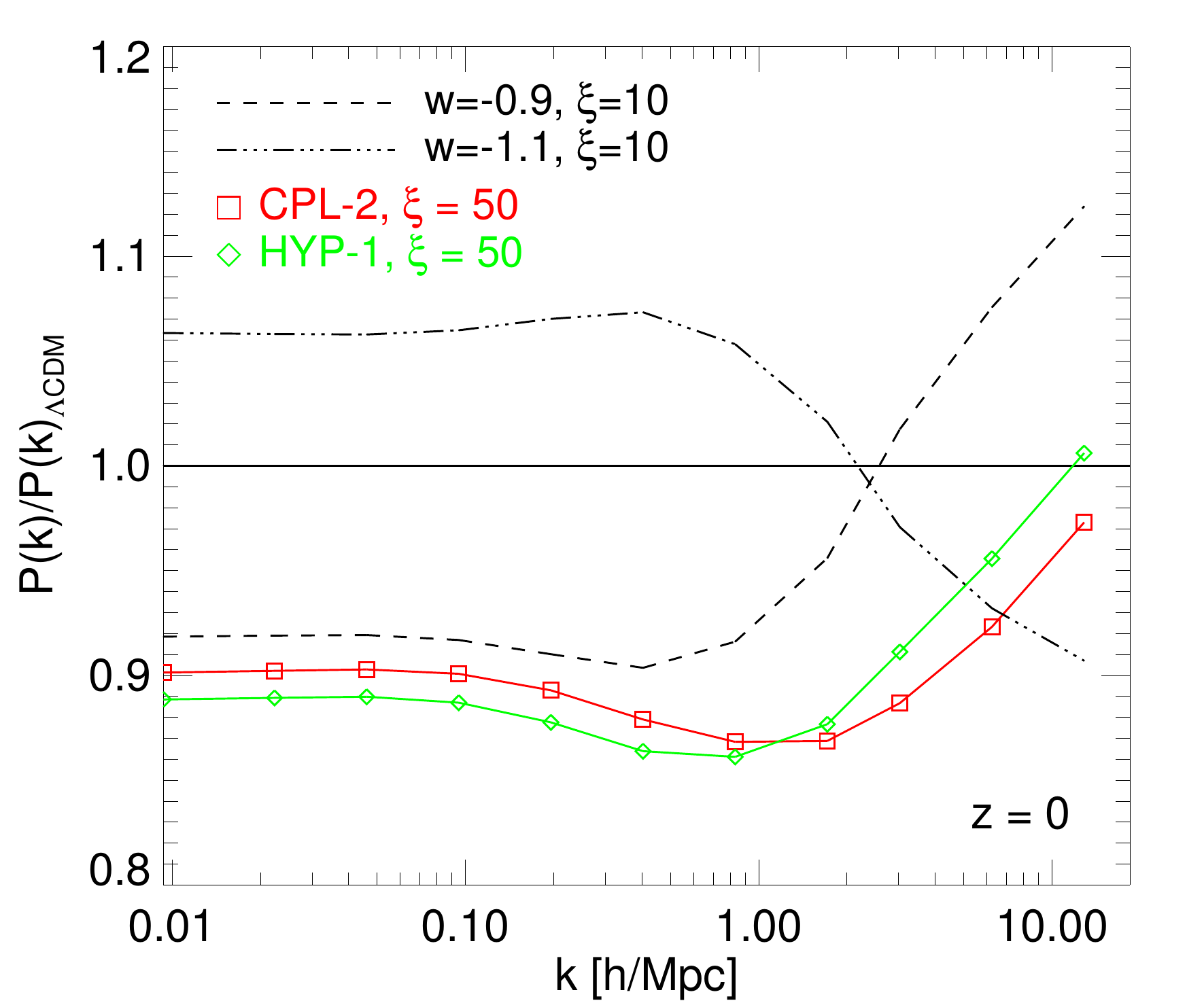}
\includegraphics[scale=0.3]{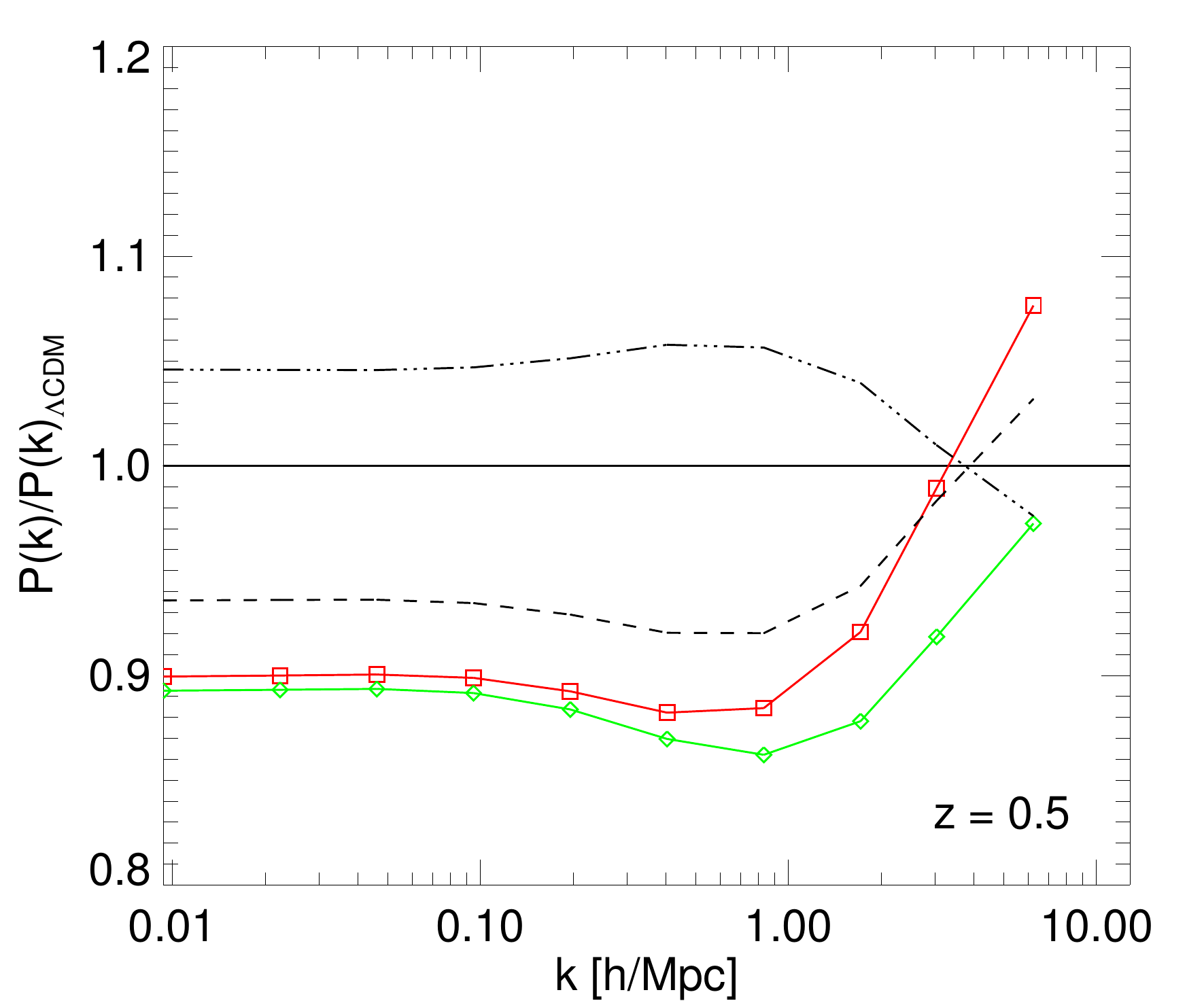}
\includegraphics[scale=0.3]{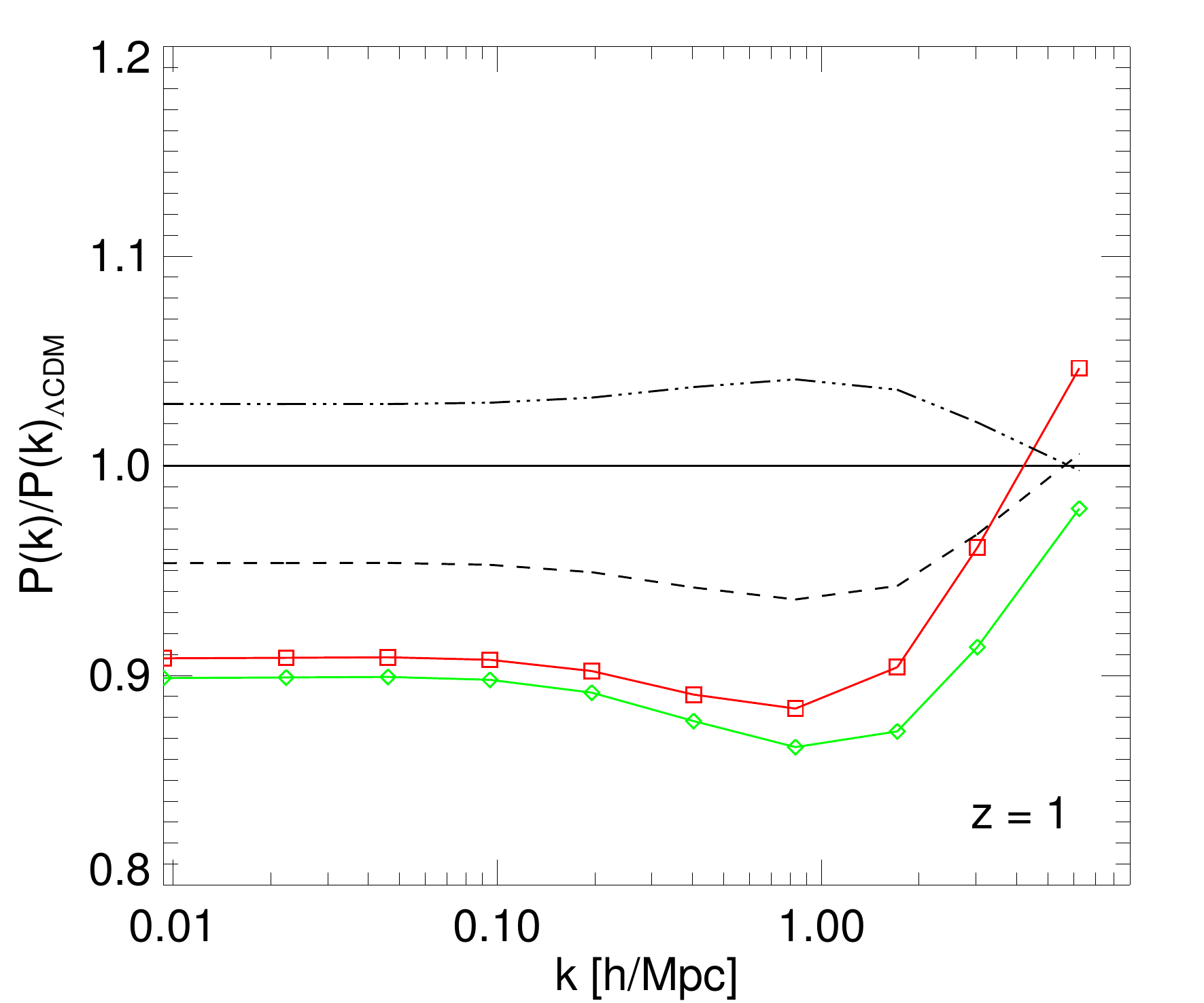}
\caption{{ The nonlinear matter power spectrum ratio to the reference $\Lambda$CDM model at three different redshifts $z=0$ ({\em left}), $z=0.5$ ({\em middle}), and $z=1$ ({\em right}) for the selected models of parameterised $w(z)$ and for the two constant-$w$ cosmologies already investigated in \citet{Baldi_Simpson_2015}. The color coding and linestyles are the same as in all previous figures.}}
\label{fig:power_ratio_LCDM}
\end{figure*}

\subsection{The Large simulations}
\label{sec:large}

We present here the results of our suite of large simulations focusing on the effects that the momentum exchange between dark energy and CDM particles has on a series of standard cosmological observables. As our analysis will show, the two selected models CPL-2 and HYP-1 will result in a cosmological evolution of structures that closely resembles that of $\Lambda $CDM for most of the observables, with the noticeable exception of the abundance of very massive objects and the overall normalisation of the linear matter power spectrum, possibly easing tensions between CMB constraints and local measurements of large-scale structures.

\subsubsection{Large-scale matter distribution}

As a first diagnostics of the effects of the momentum exchange in the two selected variable-$w$ models we compute the projected density field of a slice of thickness $30$ Mpc$/h$ through the simulation box. We assign the mass of particles in the slice to a $4096^{2}$ cartesian grid trough a Cloud-In-Cell (CIC) mass assignment scheme based on their projected positions in the $x-y$ plane and we compute the logarithm of the density contrast in the grid. In Fig.~\ref{fig:density_field} we show the density field at $z=0$ in a region of side $500$ Mpc$/h$ centered on the most massive halo identified in the simulation. In the inset displayed in the bottom-right part of the maps we show a zoom on the central halo with side $50$ Mpc$/h$.

This preliminary visual inspection shows a very similar shape of the large-scale matter distribution, with no significant difference in the location, shape, and size of overdense regions and voids. The closer look at the region around a massive halo displayed in the zoomed inset highlights an almost identical geometry of the cosmic web of filaments converging onto the central structure, even though some differences appear in the relative position of the most prominent substructures in the vicinity of the central halo.

Therefore, the overall density field seems very mildly affected by the momentum exchange with some appreciable effect showing up only in the vicinity of the most overdense regions of the simulated volume.

\begin{figure*}
\includegraphics[width=\textwidth]{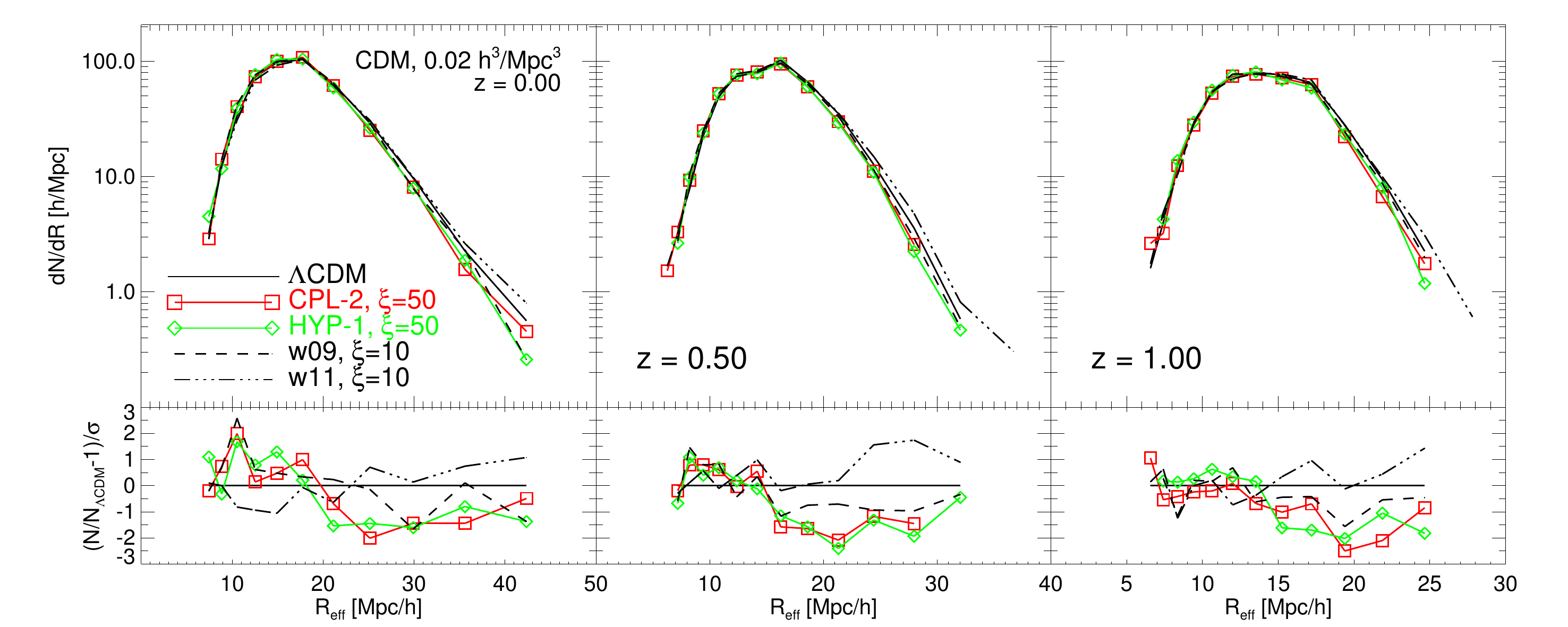}
\includegraphics[width=\textwidth]{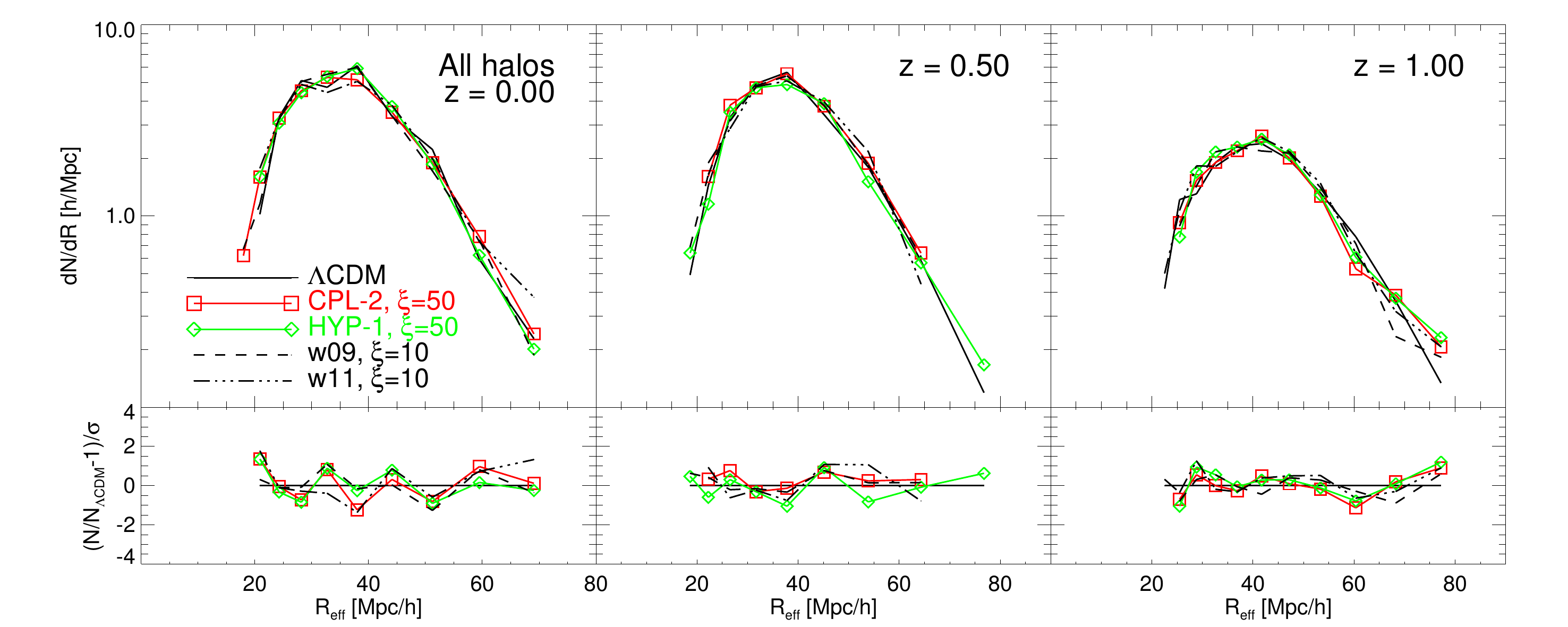}
\caption{The differential void size distribution in the CDM field (top) and in the halos distribution (bottom) as computed using the {\small VIDE} void finding toolkit. As one can see in the figures, the slight reduction in the abundance of large voids found for the CDM field is mostly erased in the distribution of halos.}
\label{fig:voids_size_dist}
\end{figure*}

\subsubsection{The nonlinear matter power spectrum}
\label{nonlinearpower}

In Fig.~\ref{fig:power_ratio_LCDM} we show the nonlinear mater power spectrum ratio to the $\Lambda $CDM reference simulation for all the models simulated in the large box. The plots are the same as in Fig.~\ref{fig:small_power_ratio_LCDM} although covering a different range of scales due to the larger size of the simulated box. As the plots clearly show, the effect of the momentum exchange at large linear scales in the CPL-2 and HYP-1 models  is twice as large as for the constant-$w$ model with $w=-0.9$ and $\xi = 10\,  [{\rm bn}/{\rm GeV}]$ at high redshift, while it becomes comparable to the latter at $z=0$. This is consistent with the variable-$w$ models having a more efficient momentum transfer at high $z$ due to the larger value of $w$ (see Fig.~\ref{fig:friction_factor}). At the same time, it is interesting to notice that also the scale dependence of the effect at small nonlinear scales is much more pronounced in the variable-$w$ models than in the constant-$w$ case at high redshift while the opposite occurs at $z=0$. This suggests that at high redshift our selected models of dark scattering might be characterized by significantly overconcentrated collapsed structures embedded in a less evolved large-scale matter distribution, as will be explicitly verified below (see Section \ref{haloconc}). Such prediction could be verified by combining weak and strong lensing observations at high $z$.

\begin{figure*}
\includegraphics[width=0.75\textwidth]{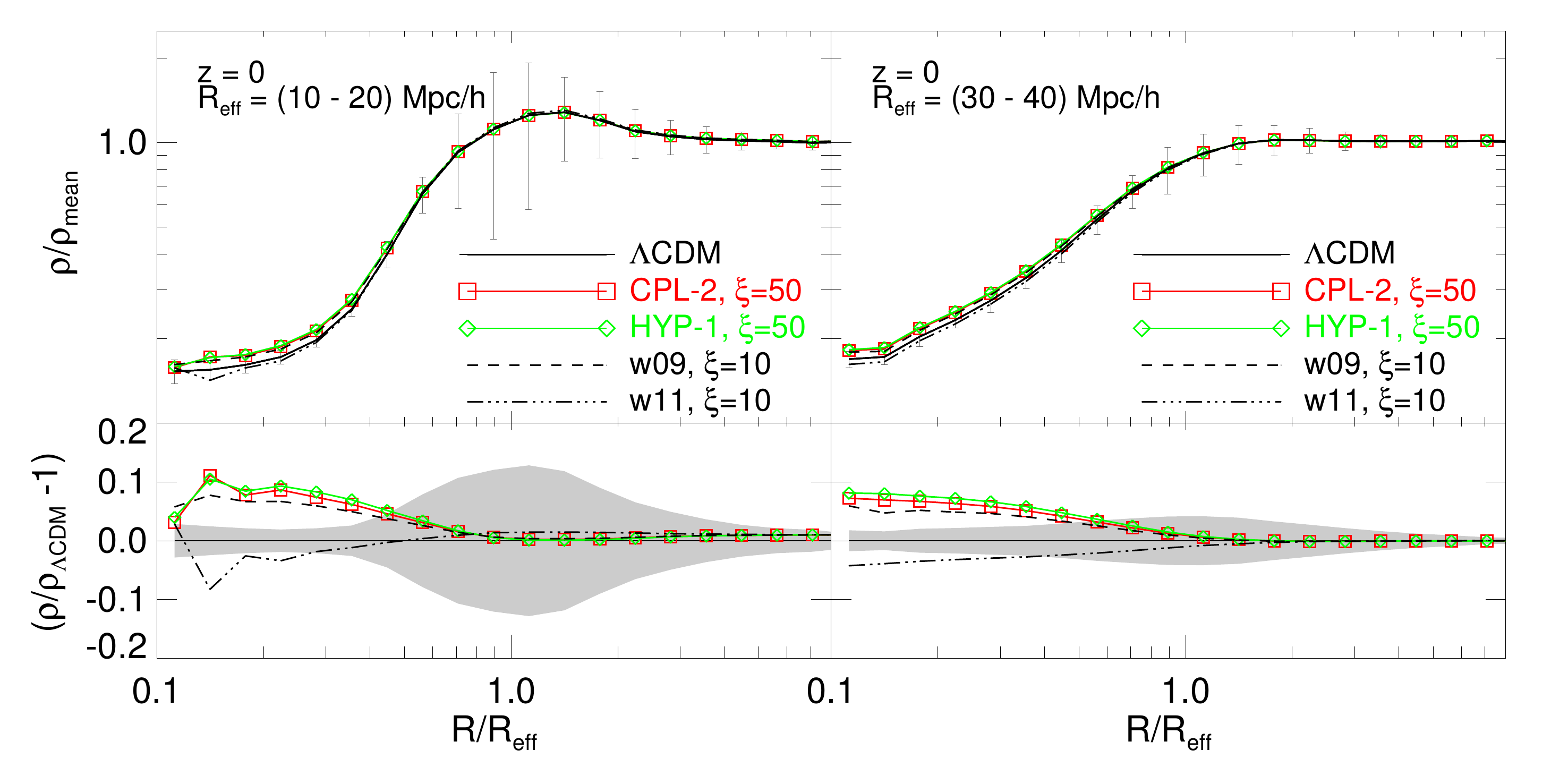}
\caption{The stacked density profiles for voids identified in the CDM distribution of the various large simulations. The stacking has been performed using $100$ randomly selected voids with effective radius in the range $10-20$ Mpc$/h$ ({\em left plot}) and $30-40$ Mpc$/h$ ({\em right plot}). The bottom panels display the ratio of the density profiles to the reference $\Lambda $CDM model, and the grey-shaded region represents the 1-$\sigma $ statistical significance according to a {\em bootstrap} estimation. As one can see in the figure, the mommentum exchange determines slightly but significantly shallower void profiles.}
\label{fig:voids_profiles}
\end{figure*}

\subsubsection{Cosmic voids}

For all our large box simulations we have identified cosmic voids using the {\small VIDE} public toolkit\footnote{https://bitbucket.org/cosmicvoids/vide\_public} \citep[][]{VIDE} in both a random subsampling of the CDM particle distribution with a tracer density of $0.02$ particles per cubic Mpc$/h$ and in the distribution of halos of our FoF sample, with a minimum halo mass $M_{FoF, {\rm min}}(z=0)\approx 2.5\times 10^{12}\, M_{\odot }/h$, for the same three redshifts investigated above $z=\left\{ 0\,, 0.5\,, 1\right\}$. The voids are identified using a Voronoi tessellation procedure and a watershed algorithm to join underdense Voronoi cells until a border to a neighboring underdense region is reached. Then, an effective radius $R_{eff}\equiv \left[ 3\cdot V_{void}/(4\pi )\right] ^{1/3}$ is associated to the void volume $V_{void}$ assuming sphericity \citep[see][for a detailed description of the algorithm implemented in the VIDE code]{VIDE}. 

Starting from the void catalogue produced by {\small VIDE} we identify the main voids (i.e. those voids that are not embedded within larger voids) and remove pathological voids following the procedure described in \citet{Pollina_etal_2016}. In this way, we ensure that the final void catalogue contains only disjoint voids with a central overdensity $\delta _{\rm min} < 0.2$ and a density contrast between the density minimum and the void boundary $\delta _{\rm c} > 1.57$. With such catalogue of selected voids we compute the differential void size  distribution, i.e. the number density of voids as a function of their effective radius $R_{eff}$. The comparison of the void size distribution functions of the different models is displayed in Fig.~\ref{fig:voids_size_dist} where the upper plot refers to the voids in the subsampled CDM field while the lower plot to the voids in the FoF halos catalogs.
In each plot the top panel shows the void size distribution function while the bottom panel presents the relative difference with respect to the $\Lambda $CDM reference in units of the statistical significance computed by propagating the Poisson noise in each bin of effective radius to the relative difference.

\begin{figure*}
\includegraphics[width=0.3\textwidth]{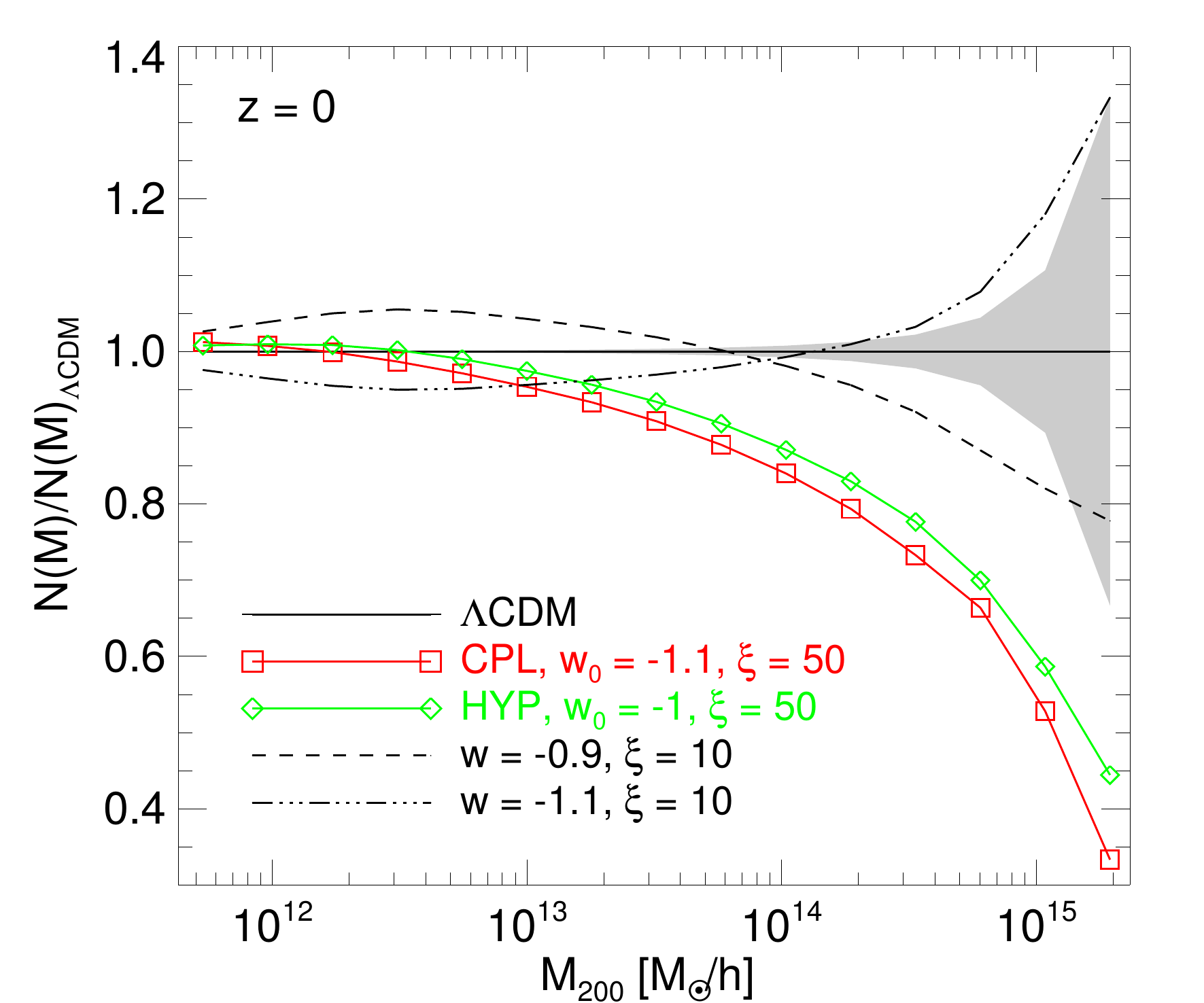}
\includegraphics[width=0.3\textwidth]{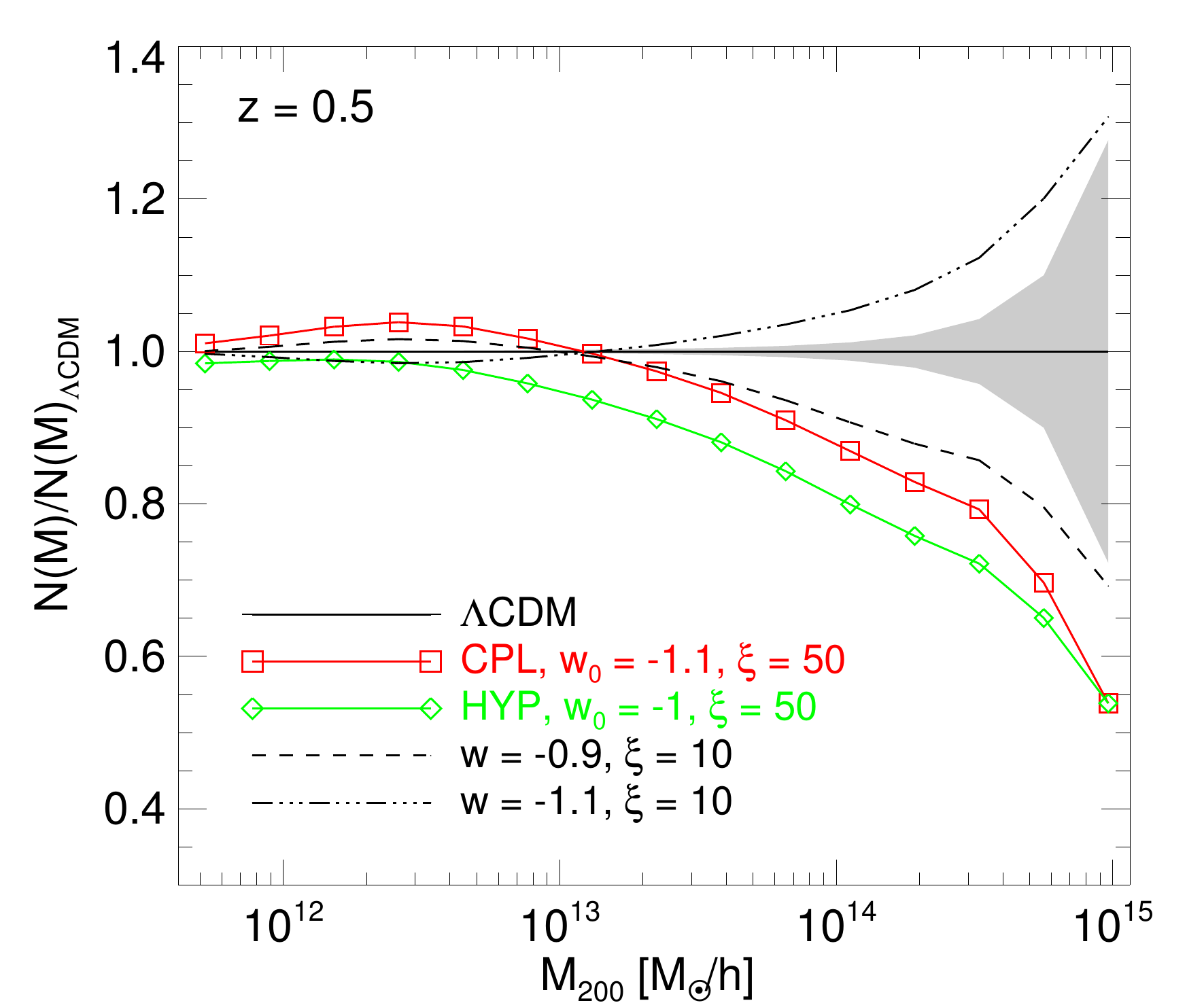}
\includegraphics[width=0.3\textwidth]{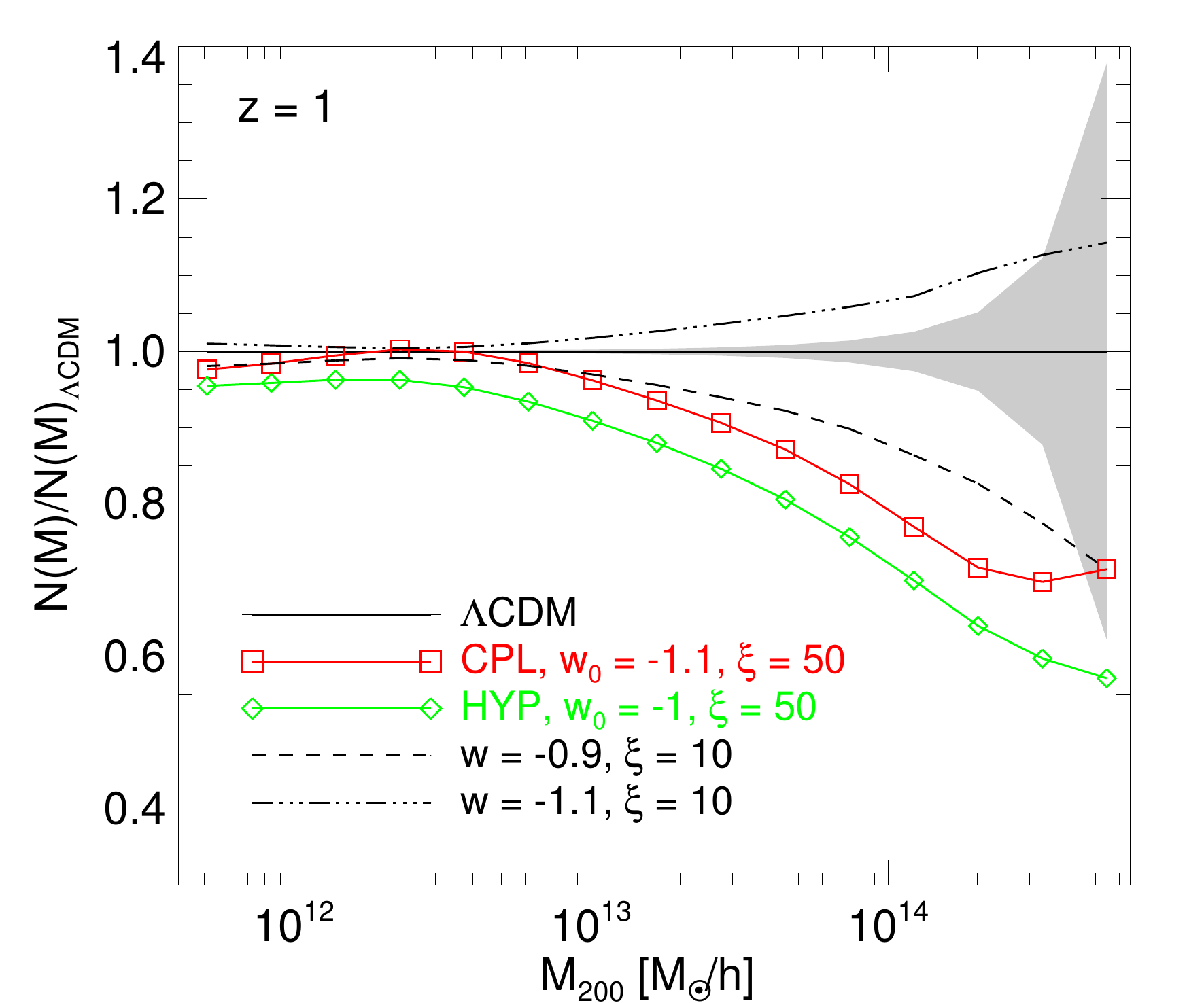}
\caption{{ The mass function ratio to the $\Lambda $CDM case for the models included in our suite of large-svale simulations. The three redshifts displayed in the different panels, as well as colours, symbols, and line styles are the same as displayed in Fig.~\ref{fig:power_ratio_LCDM}. As one can see in the figures, the variable-$w$ models -- differently from the constant-$w$ cases -- determine a very significant suppression of the abundance of cluster-sized halos, thereby alleviating current persisting tensions between low-$z$ observational data and best-fit cosmological parameters derived from primary CMB anisotropies. }}
\label{fig:HMF_ratio_LCDM}
\end{figure*}

As one can see from the plots, the variable-$w$ models CPL-2 and HYP-1 show a slight suppression of the abundance of large voids in the CDM density field compared to the reference $\Lambda $CDM simulation. The effect is not too dramatic, with a significance level of $\approx 1-2\sigma $, and is consistent with the suppression of large scales perturbations already shown by the power spectrum comparison. Interestingly, also for this class of models -- as it has already been shown to occur for modified gravity \citep[][]{Cai_etal_2014,Achitouv_etal_2016}, massive neutrinos \citep[][]{Massara_Villaescusa-Navarro_Viel_2015} and interacting dark energy \citep[][]{Pollina_etal_2016} cosmologies -- this effect is strongly suppressed when looking at the distribution of voids in the halo catalogs, where no significant deviation can be observed besides statistical oscillations around the reference model, at all redshifts.

We also compared the structural properties of voids in the different models by computing the stacked radial density profiles around the voids macrocenters identified in the $\Lambda $CDM simulation for 100 randomly selected voids within two bins of void effective radius, namely $R_{eff}\in \left\{10-20\,, 30-40\right\}$ Mpc$/h$. The results are shown in Fig.~\ref{fig:voids_profiles} and clearly show how the variable-$w$ models of momentum exchange result in shallower profiles for voids of both  bins. Therefore, voids appear to be less empty in Dark Scattering cosmologies, which might result in a lower weak lensing signal at low multipoles.

{
\begin{figure*}
\includegraphics[width=0.3\textwidth]{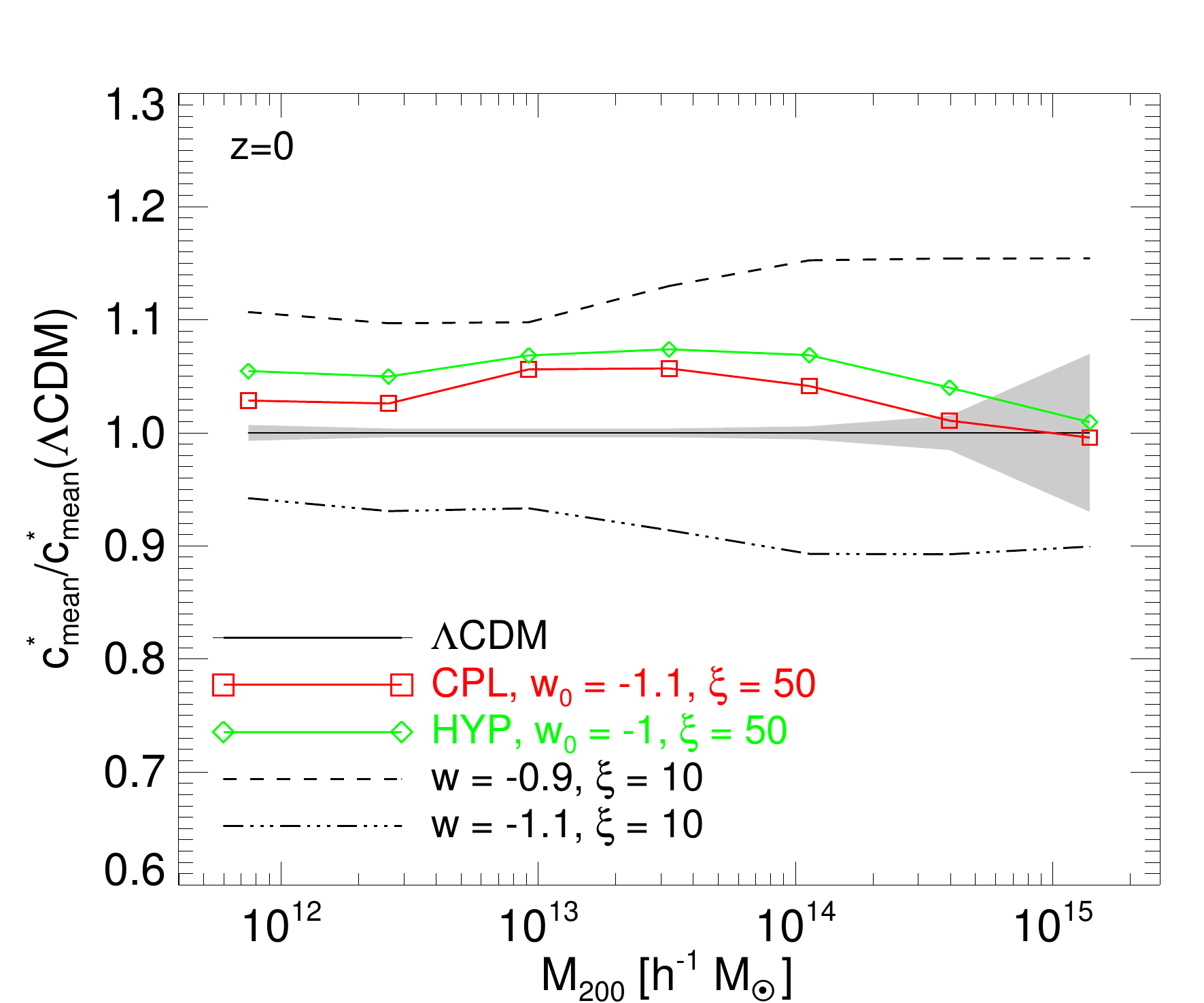}
\includegraphics[width=0.3\textwidth ]{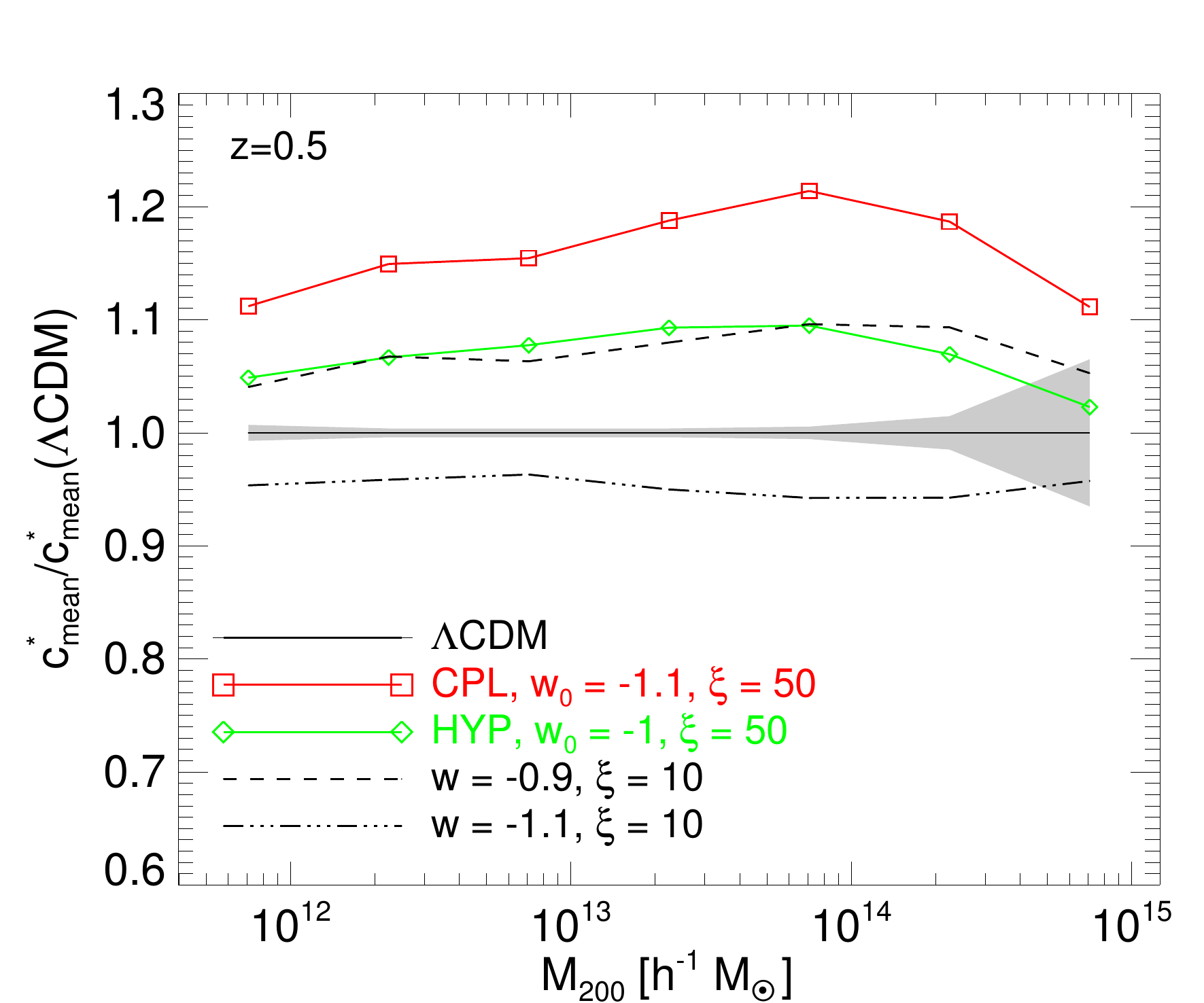}
\includegraphics[width=0.3\textwidth]{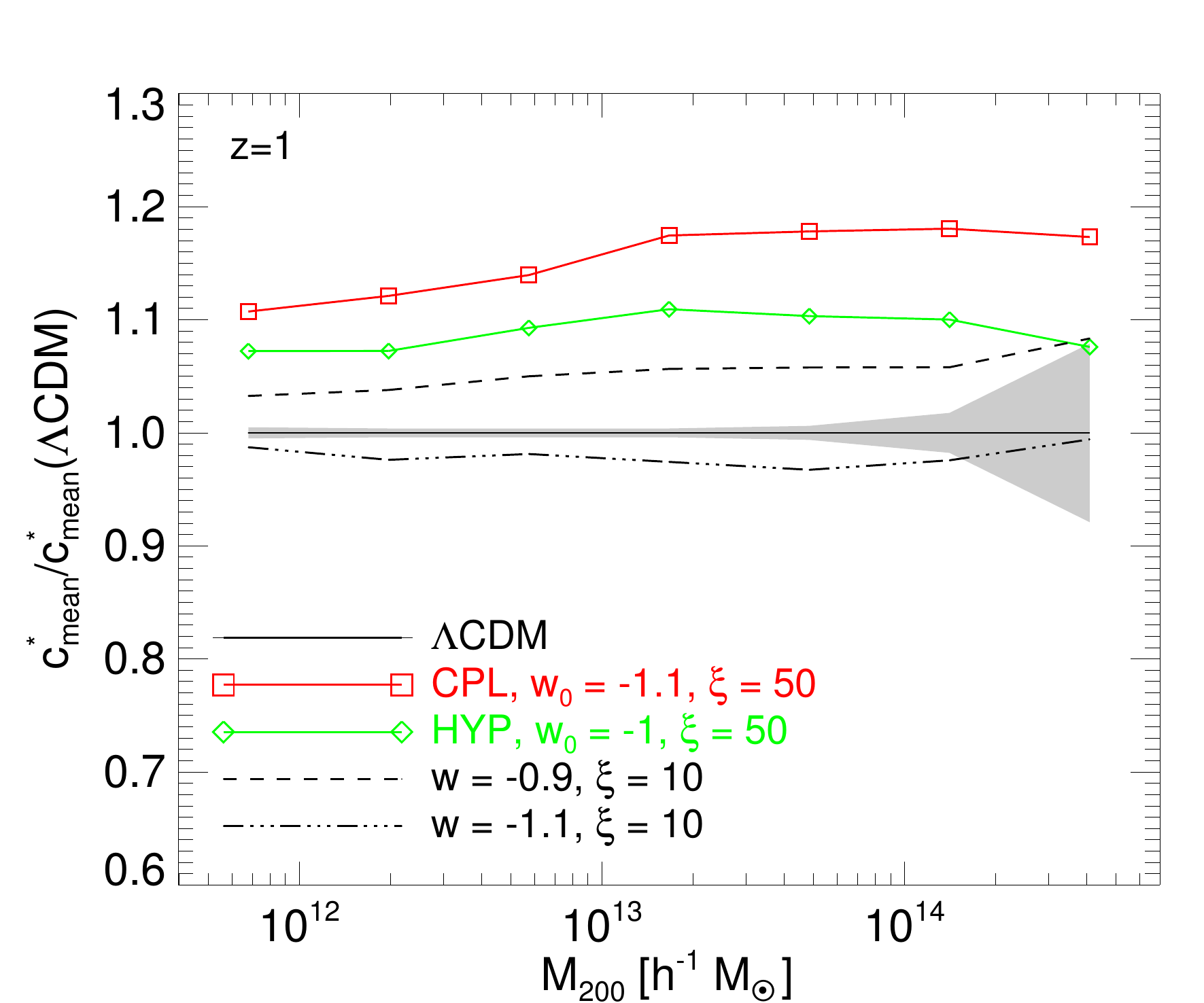}
\caption{{The ratio of the binned average concentration to the $\Lambda $CDM reference at the same three different redshifts considered in the previous figures. As one can see in the plots, the variable-$w$ models do not show a strong impact on the concentrations of halos at very low redshifts, while at higher redshifts the effect appears to be somewhat enhanced.}}
\label{fig:concentrations}
\end{figure*}
}

\subsubsection{The halo mass function}

As already done in Fig.~\ref{fig:small_HMF_ratio_LCDM} for the intermediate-scale simulations, in Fig.~\ref{fig:HMF_ratio_LCDM} we display the ratio of the differential halo mass function to the $\Lambda $CDM case for the models that were simulated in the larger boxes. This allows to increase the statistics of massive halos and extend the range of the computed mass function to larger masses, thereby investigating the impact of the momentum exchange on the abundance of massive clusters of galaxies as a function of redshift.

This more extended mass range allows to see that both the variable-$w$ cosmologies have a very significant impact on the abundance of very massive objects, suppressing the number density of clusters with mass around $10^{15}h^{-1}$ M$_{\odot }$ by $\approx 40-50\%$ at $z=0$. The effect is milder at higher redshifts, but still significant with a suppression of the abundance of $10^{14}h^{-1}$ M$_{\odot }$ halos by $\approx 20-30\%$ at $z=1$.

This represents one of the most prominent features of the cosmological models under investigation and might provide a way to reconcile the low abundance of SZ clusters detected by Planck \citep[][]{Planck_XXIV} with their expected number based on the cosmological constraints arising from the angular power spectrum of temperature and polarisation anisotropies \citep[][]{Planck_2015_XIII}.
 It is also remarkable that the impact on the abundance of smaller objects -- down to Milky-Way sized halos of $\sim 10^{12}$ M$_{\odot }/h$ -- is very mild, thereby leaving unaffected the expected number of galaxies.

\begin{figure*}
\includegraphics[scale=0.44]{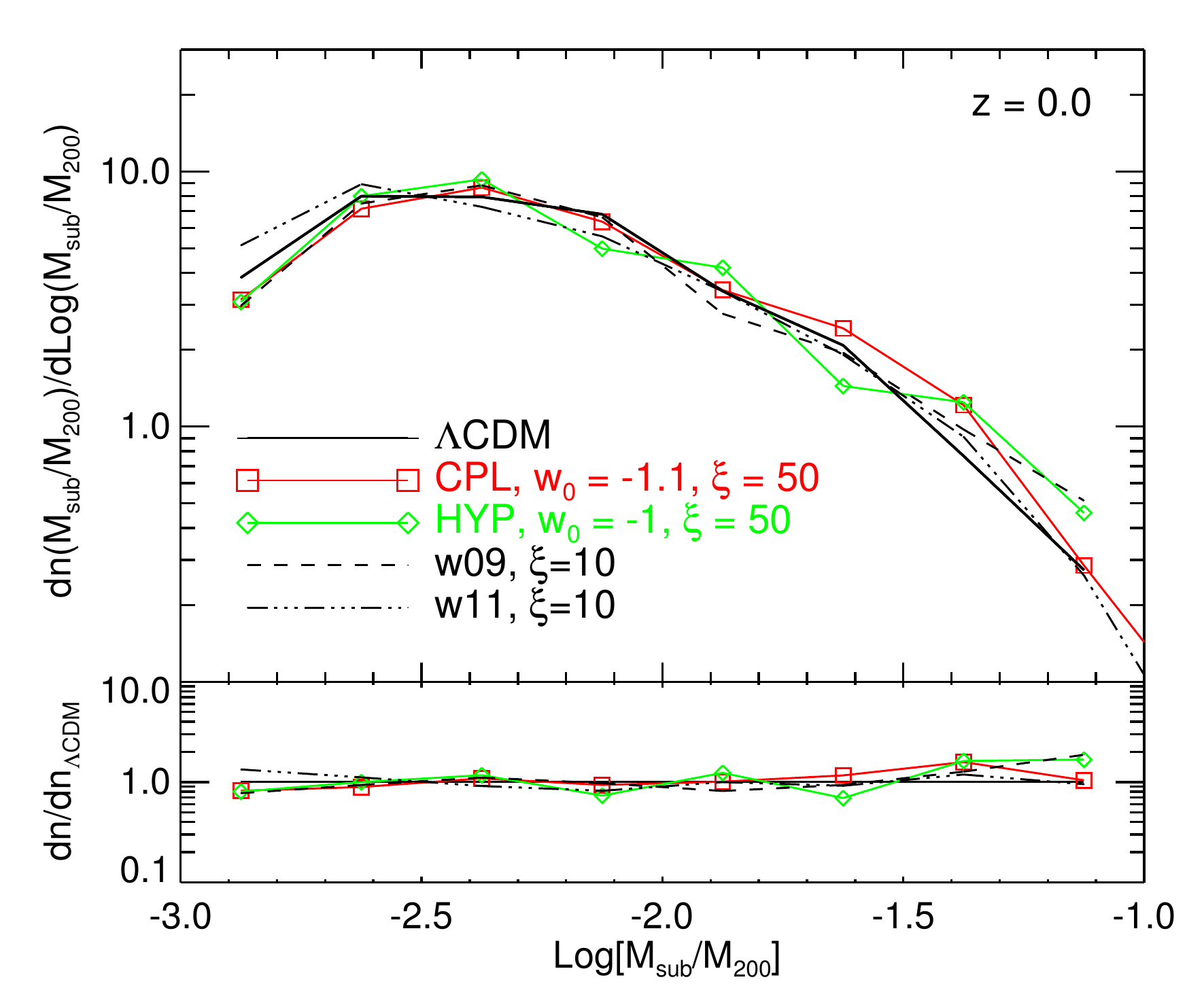}
\includegraphics[scale=0.44]{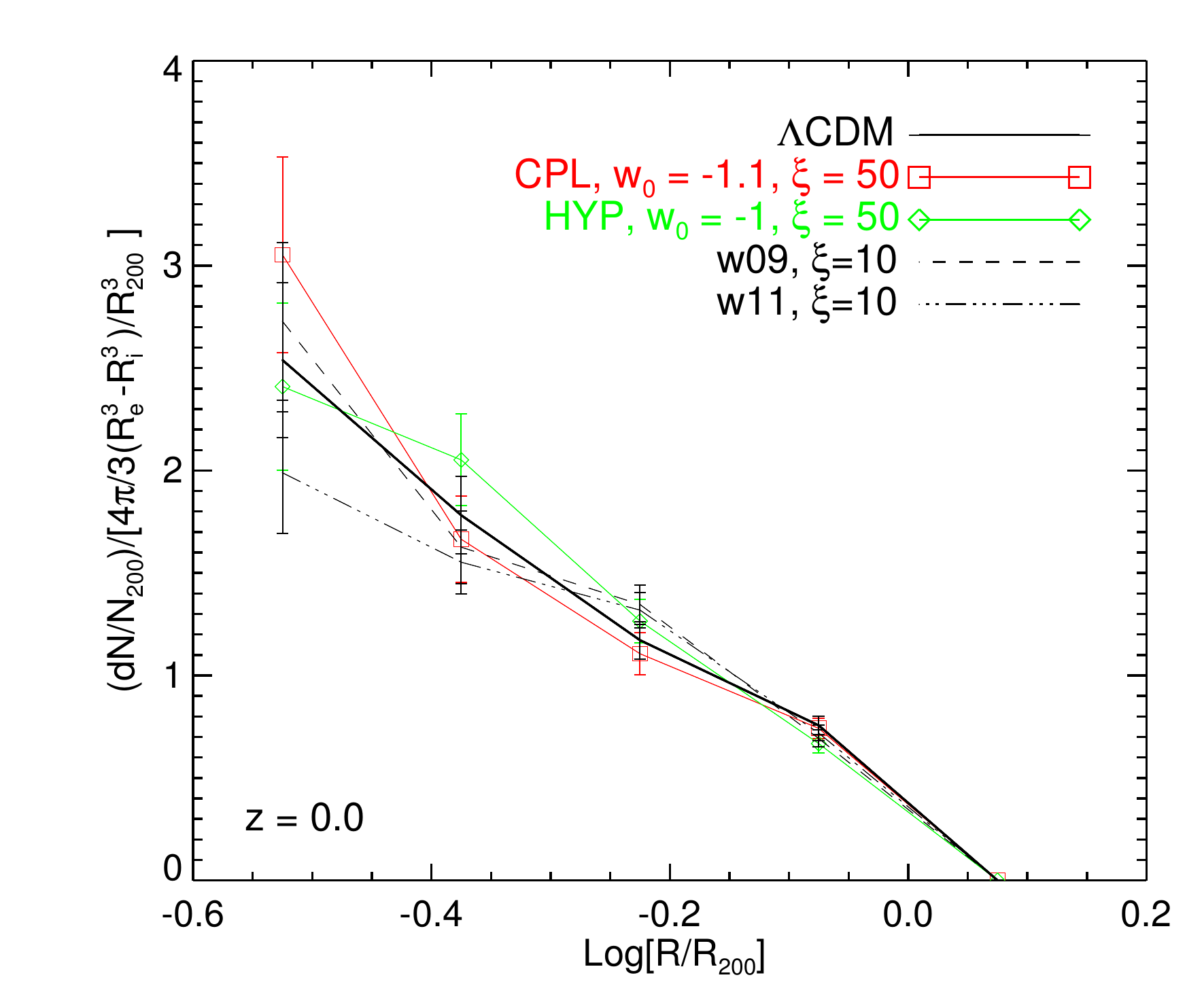}
\caption{{ The sub halo mass function ({\em left}) and the sub halo radial distribution ({\em right}) for the various models considered in our suite of large-scale simulations at $z=0$. As one can see in the plots, the momentum exchange does not significantly alter the abundance and the radial distribution of substructures identified in the mass range allowed by the resolution of our simulations.}}
\label{fig:subhalo_mass_function}
\end{figure*}

\subsubsection{Halos concentrations}
\label{haloconc}

For each halo in our sample we have computed the concentrations $c^{*}$ following the approach described in \citet{Aquarius} as:
\begin{equation}
\frac{200}{3}\frac{c^{*3}}{\ln (1+c^{*}) - c^{*}/(1+c^{*})} = 7.213~\delta _{V}
\end{equation}
where $\delta _{V}$ is defined as:
\begin{equation}
\delta _{V} = 2\left( \frac{V_{max}}{H_{0}r_{max}}\right) ^{2}
\end{equation}
with $V_{max}$ and $r_{max}$ being the maximum rotational velocity of the halo and the radius at which this velocity peak is located, respectively. 

In Fig.~\ref{fig:concentrations} we show the ratio of these average concentrations at the usual three different redshifts $z=\left\{ 0, 0.5,1\right\}$ as a function of the halo mass $M_{200}$ for a set of logarithmically equispaced mass bins. The grey shaded area shows the Poissonian error based on the number of halos in each bin for the reference $\Lambda $CDM run.
As the figures show, at $z=0$ the variable-$w$ models have a significantly weaker impact on the halo concentrations than the constant-$w$ ones even for a higher value of the interaction parameter $\xi$. Both models determine an increase of concentrations below $7\%$ over the whole mass range covered by our halo catalogs. This implies that no dramatic effect in low-redshift strong lensing observations is expected for the models under investigation.  On the other hand, as anticipated in Section \ref{nonlinearpower}, the situation changes at higher redshifts where the models with a variable equation of state show a more significant increase of halo concentrations. {They are comparable to (at $z=0.5$) or even larger than (at $z=1$) the constant-$w$ case} (even though the latter  {has a weaker} interaction parameter $\xi $). Therefore, these models predict a somewhat enhanced strong lensing efficiency at high redshift despite the overall reduction of large-scale power which is expected to result in a lower weak lensing signal.

\subsubsection{Halo substructures}

As a final probe of the momentum exchange between dark energy and dark matter particles we investigate the abundance and spatial distribution of substructures within massive collapsed halos. 

First, we compute the subhalo mass function, defined as the number of substructures with a given fractional mass with respect to the virial mass of their host main halo ($M_{\rm sub}/M_{200}$) as a function of the fractional mass itself. We compute this quantity by binning in logarithmic fractional mass bins the whole sample of substructures belonging to host halos with virial mass above a minimum threshold of $M_{\rm min}=10^{14}$ M$_{\odot }/h$, thereby restricting this analysis to massive galaxy cluster halos (due to the limited resolution of the simulations). 
In the left panel of Fig.~\ref{fig:subhalo_mass_function} we show the subhalo mass function at $z=0$ for the various models and in the bottom plot we display their ratio to the fiducial $\Lambda $CDM cosmology. 

Then, we also compute the subhalo radial distribution (shown in the {\em Right} panel of Fig.~\ref{fig:subhalo_mass_function}), defined as the fractional number density of substructures in a series of logarithmically equispaced radial bins in units of the virial radius $R_{200}$ of the host halo. 

Both these observables show very little deviations among all the models and the reference $\Lambda $CDM cosmology, thereby showing that the momentum exchange does not significantly alter the distribution of substructures at small scales.

\subsubsection{Halo velocity dispersions}

\begin{figure}
\includegraphics[scale=0.44]{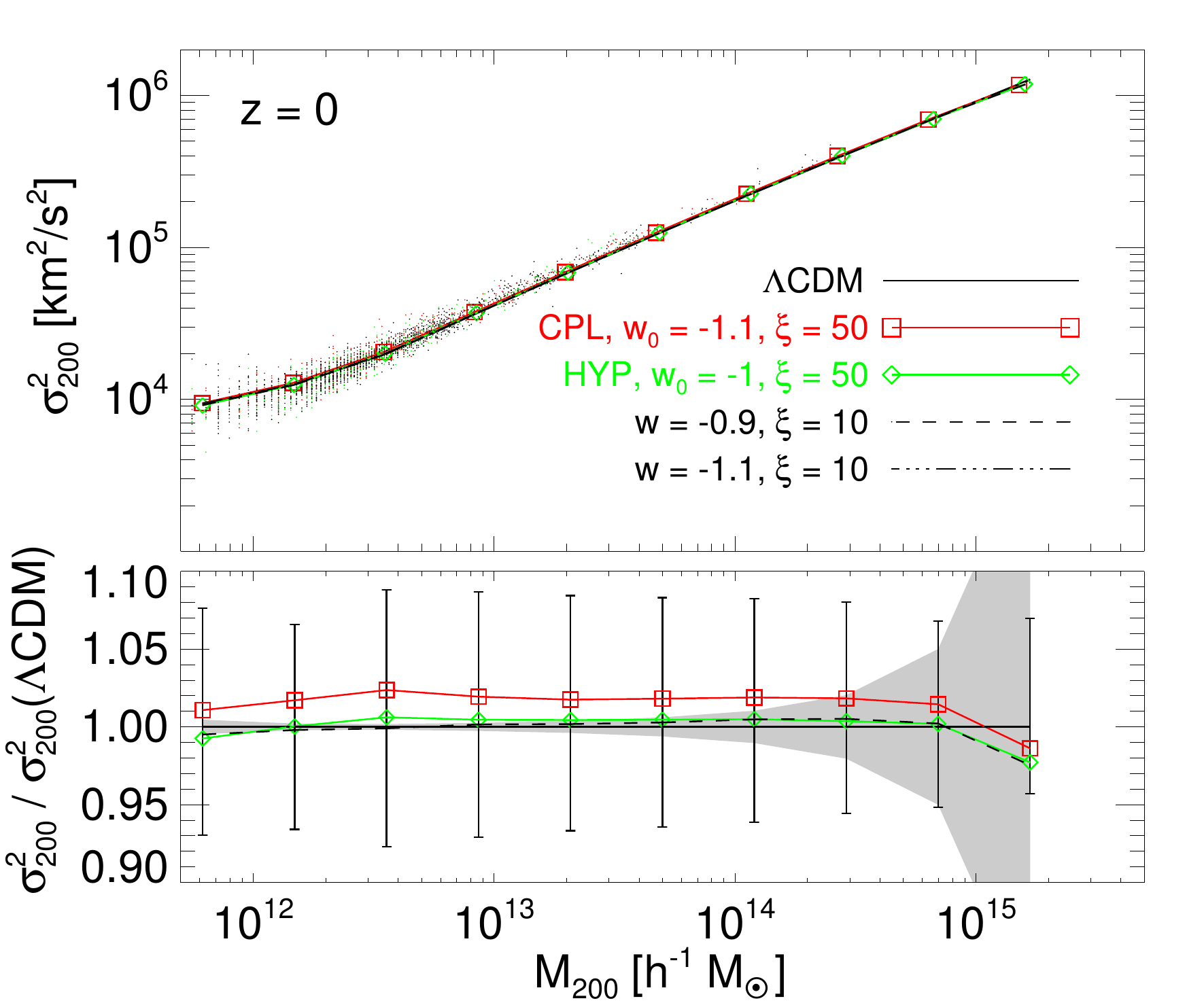}
\caption{The velocity dispersion of halos at $z=0$ as a function of halo mass for the various models under investigation. The overall deviation from the reference $\Lambda $CDM cosmology does not exceed a few percent over the whole mass range accessible to our simulations.}
\label{fig:veldisp}
\end{figure}

For all the halos of each simulation sample we compute the one-dimensional velocity dispersion $\sigma ^{2}$ and compare it to the corresponding behavior of the $\Lambda $CDM run. {The results are shown in Fig.~\ref{fig:veldisp} for the present epoch ($z=0$),  across 10 logarithmically equispaced mass bins for the different models. In the upper panel we display as coloured dots a random subsample of all the halos in the catalogs, while  the lines trace the mean value of $\sigma ^{2}$. }
% Above sentence was a bit long
In the bottom panels we plot the ratio of the binned average 1-D velocity dispersion to the $\Lambda $CDM case, and the grey shaded region indicates the Poissonian error associated with number counts of halos in each bin of the reference simulation. 

As one can see from the plots, the Dark Scattering models generate a very mild enhancement (only $\approx 1-2\%$) of the 1-D velocity dispersion with respect to the $\Lambda $CDM reference over the whole mass range covered by our simulations, with a roughly mass-independent behaviour.

\subsubsection{Halo bias}

\begin{figure}
\includegraphics[scale=0.44]{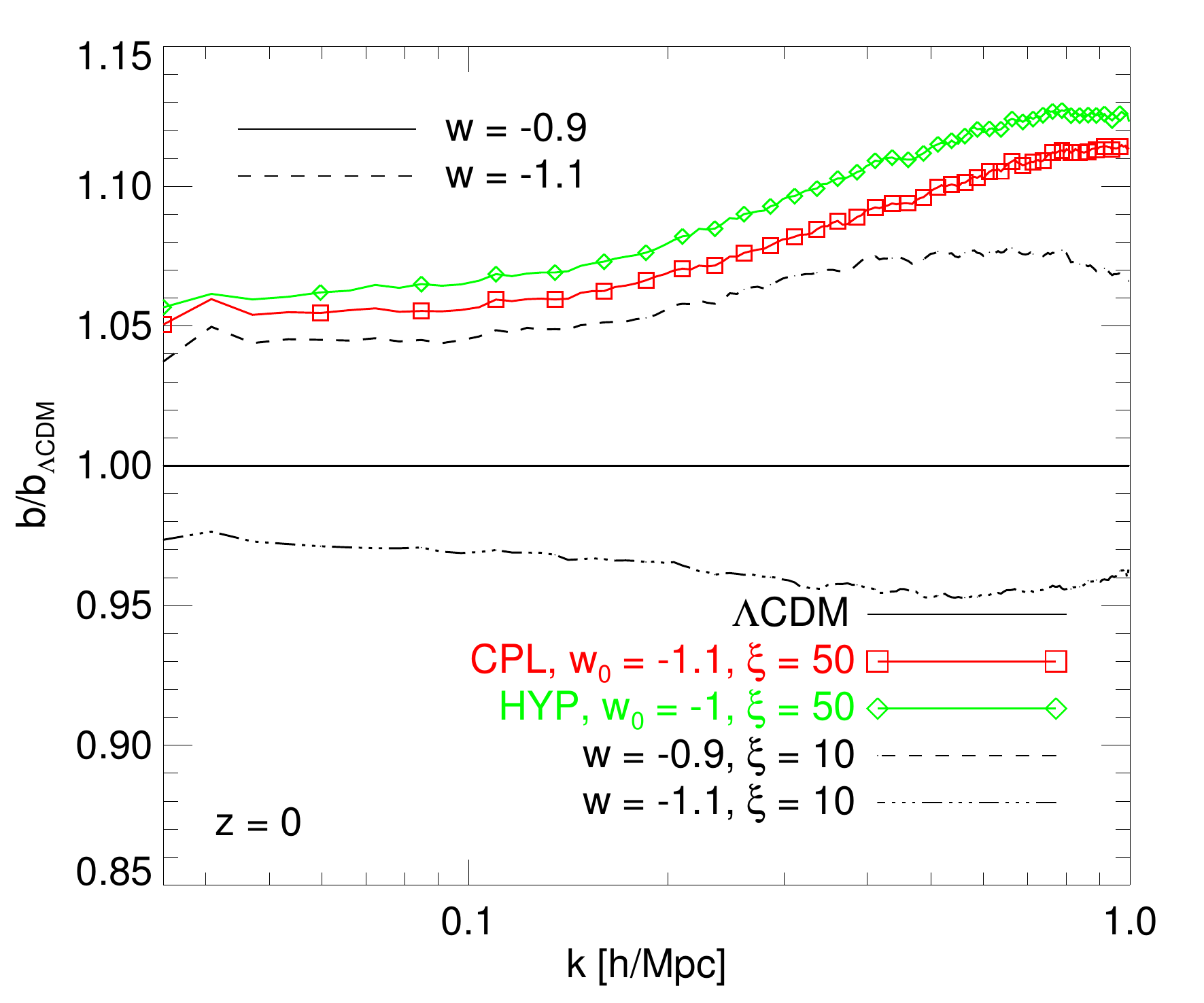}
\caption{The halo bias as a function of scale at $z=0$ for the models considered in the large-scale simulations. As one can see in the figure, the momentum exchange determines a slight increase of the bias with a characteristic scale-dependence of the enhancement.}
\label{fig:bias}
\end{figure}

As a final test of our models we compute the halo bias by taking the ratio of the halo-matter cross-power spectrum $P_{\rm hm}(k)$  to the matter-matter power spectrum $P_{\rm mm}(k)$ for halos with mass above $5\times 10^{11}$ M$_{\odot }/h$. In Fig.~\ref{fig:bias} we display the ratio of the halo bias of the various models to the $\Lambda $CDM case in the range of scales $0.04\leq k  \cdot h/{\rm Mpc} \leq 1$. 
As already shown in \citet{Baldi_Simpson_2015}, the constant-$w$ models determine a $\approx 5-7\%$ increase (reduction) of the halo bias with a slight scale-dependence for the interaction parameter $\xi = 10$ adopted in our simulations for $w=-0.9$ ($w=-1.1$). 

Interestingly, the variable-$w$ models under investigation, despite the higher interaction parameter $\xi = 50$, are found to {yield only a marginal increase of the large scale bias. The effect at the smallest scales under consideration is moderately stronger than before,} reaching a $\approx 10-12\%$ deviation from $\Lambda $CDM. Furthermore, these models show a more pronounced scale-dependence, which might represent a further testable prediction of the Dark Scattering scenario.

\section{Conclusions}
\label{sec:concl}

The physical nature of dark matter and dark energy remains highly active topic of theoretical and experimental investigation.   In this work we began by highlighting a connection between the phenomenologically motivated elastic scattering model, and a subclass of coupled scalar field models. Furthermore we showed that, quite generically, the prediction of models which invoke momentum exchange between dark energy and dark matter is one where the linear growth rate is modulated in a scale-independent manner. In order to investigate further consequences of this interaction, we performed a series of N-body simulations, building on the work of \citet{Baldi_Simpson_2015}, but now incorporating a more realistic trajectory for the dark energy equation of state. In this context, the key consequence of an evolving equation of state - specifically freezing models which tend to mimic a cosmological constant at late times - is that the coupling is naturally weakened in the late universe, when the bulk of the non-linear structure formation takes place. As a result, the amplification of the non-linear matter power spectrum which had been found in the earlier simulations of \cite{Baldi_Simpson_2015} is significantly suppressed. \\

{We began by considering two different realisations of a freezing Dark Energy equation of state $w(z)$ with either negative or positive convexity near the present epoch, namely the widely-used CPL parameterisation and a novel step-like parameterisation based on hyperbolic functions, respectively. We then explored the relevant parameter space of these two forms of freezing Dark Energy models with a suite of intermediate-size simulations, assuming a relatively large value of the interaction parameter ($\xi = 50\,{\rm bn}/{\rm GeV}$). With these simulations at hand, we identified those specific models that appear to affect some basic statistics of the large-scale matter distribution (such as the nonlinear matter power spectrum and the halo mass function) in the direction of a suppressed growth of perturbations, without impacting too dramatically on the highly nonlinear regime at very small scales. For such models, we performed simulations on larger scales to improve the statistics of our results.\\

Finally, by means of these larger simulations, we {have been able to verify} that a number of observable quantities are not significantly perturbed by the interaction.   More specifically:
\vspace{-0.18 cm}
\begin{itemize}
\item[$\star $] The density field of the Dark Scattering models shows the same shape of the large-scale cosmic web as compared to the standard $\Lambda $CDM realisation, with tiny differences in the position of prominent substructures around very massive systems appearing only at very small scales;
\item[$\star $] The abundance of galaxy-sized and group-sized halos (up to $10^{13}$ M$_{\odot }/h$) is mildly affected by the interaction in the redshift range $0\leq z \leq 1$;
\item[$\star $] The abundance of cosmic voids identified in the distribution of collapsed halos does not show any significant deviation compared to $\Lambda $CDM;
\item[$\star $] The relative abundance of substructures encoded by the subhalo mass function as well as the radial distribution of substructures around their host halo is basically unchanged, with differences appearing consistent with statistical uncertainties.
\item[$\star $] The 1-dimensional velocity dispersion of halos is only very slightly enhanced in Dark Scattering models as compared to $\Lambda $CDM, with an effect not exceeding $\approx 1-2\%$ at $z=0$;
\item[$\star $] The halo bias is weakly increased (in a scale-dependent fashion), with a deviation from the $\Lambda $CDM reference ranging between $\approx 5\%$ at large scales ($k\approx 10^{-2}\, h/$Mpc) and $\approx 12\%$ at small scales ($k\approx 1\, h/$Mpc); more specifically, the bias at $k\sim 0.05-0.1\, h/$Mpc is enhanced from the value $b\approx 1.08$ in $\Lambda $CDM to $b\approx 1.15$ for both the CPL and HYP parameterisations.
\end{itemize}
On the other hand, we identified a set of characteristic observational footprints of the model that might alleviate the persisting tensions between high-redshift and low-redshift cosmological constraints. In particular:
\begin{itemize}
\item[$\star $] The matter power spectrum is suppressed at linear scales by up to $\sim 10\%$ at $z=0$ for the specific parameters considered in our analysis, thereby providing a weaker contribution to the overall weak lensing signal and a lower determination of $\sigma _{8}$ from low-redshift probes;
\item[$\star $] The abundance of cluster-sized halos is very significantly reduced, with a suppression reaching $\sim 60\%$ for halos of $\sim 10^{15}$ M$_{\odot }/h$ at $z=0$;
\item[$\star $] The concentration-mass relation, which appears very mildly affected at the present epoch (with a mass-independent increase of the normalisation below $7\, \%$) shows a somewhat larger deviation from $\Lambda $CDM at higher redshifts, with an enhancement up to $\sim 20\%$ at $0.5\leq z \leq 1$, which might result in a higher efficiency for strong lensing at these redshifts;
\item[$\star $] The abundance of cosmic voids in the CDM distribution shows a statistically significant suppression (at the level of about $1-2\sigma $) for voids with radius $\gtrsim 20$ Mpc$/h$, while the spherically-averaged stacked density profiles of voids are found to be shallower, with a $\approx 10\%$ higher central overdensity compared to $\Lambda $CDM, which is also consistent with a lower expected weak lensing efficiency;
\end{itemize}
 
% - such as the void size distribution, the concentration-mass relation of halos, and the sub-halo mass function. The most substantial modification to clustering statistics is the suppression of the  linear growth rate, leading to a $10\%$ reduction in linear power at low redshift.  This is also associated with a substantial reduction in the abundance of the most massive halos, those associated with galaxy clusters. 

To summarise, the cosmological models featuring elastic scattering between CDM particles and a Dark Energy field with variable equation of state $w(z)$ defined as in Eqs.\ref{CPL} and \ref{HYP} are {challenging to distinguish} from the standard $\Lambda $CDM cosmology at the level of the background expansion history, while providing a growth of structures that results in a lower amplitude of linear density perturbations (i.e. lower weak lensing signal, lower $\sigma _{8}$) and a strongly suppressed abundance of very massive clusters at redshifts below 1. A similar conclusion has been recently reached -- based on linear structure formation only -- for some specific realisations of {\em Type 3} interacting Dark Energy models by \citet{Pourtsidou_Tram_2016}. In this respect, these models might alleviate current observational tensions and deserve further investigations.\\

As a final note, let us briefly consider the relative merits of interpreting abnormal structure formation as either signs of coupled dark energy or as an indication of modified gravity.  A priori, before any experiments are conducted, we might consider both scenarios equally likely. But over the past few decades, a huge swathe of the modified gravity landscape has been erased. Following high precision measurements of local gravitational potentials within the solar system, and cosmological potentials in the cosmic microwave background, only a small fraction of (the very large) parameter space remains viable. A further null result was recently derived from the velocity of gravitational waves in the LIGO detection of merging black holes \cite{LIGO_detection}. So there have been numerous opportunities for modified gravity to show its face elsewhere, but it has failed to. In contrast, coupled models do not directly alter the motions of planets, the gravitational potentials, nor the speed of gravitational waves. Crucially then, from the perspective of Bayesian model selection, coupling models are favoured because they are more predictive. They induce a pure change in cosmological structure formation, while naturally satisfying these complementary tests. Furthermore, {a broad spectrum of} modified gravity models generically predict an enhancement of structure formation, which would aggravate rather than alleviate the aforementioned tensions in observational data.
}\\

Ultimately these two distinct explanations are observationally distinguishable, due to the coupled model producing a small segregation of the baryonic and dark matter motions. This would represent a highly challenging but not insurmountable task for future cosmological surveys.

\section*{Acknowledgments}

We are deeply thankful to Alkistis Pourtsidou for useful discussions on the connection between the Dark Scattering models and the Type 3 class of coupled dark energy. {FS would like to thank Baojiu Li for helpful comments.}
MB acknowledges support from the Italian Ministry for Education, University and Research (MIUR)
through the SIR individual grant SIMCODE, project number RBSI14P4IH.
{FS acknowledges support by the European Research Council under the European Community's Seventh Framework Programme FP7-IDEAS-Phys.LSS 240117. }
The numerical simulations presented in this work have been performed 
and analysed on the Hydra cluster at the RZG supercomputing centre in Garching.

%*****************************************************************************
\bibliographystyle{mnras}
\bibliography{dm_scatter.bib,baldi_bibliography.bib}
\label{lastpage}

\end{document}